\documentclass[twoside,8pt]{article}
\usepackage{epsfig}

\newcommand{\be}{\begin{equation}}
\newcommand{\ee}{\end{equation}}
\newcommand{\bea}{\begin{eqnarray}}
\newcommand{\eea}{\end{eqnarray}}

\begin{document}

\title{Effects of the nuclear equation of state on  the r-mode instability and evolution of neutron stars}

\author{Ch.C. Moustakidis$^{1,2}$\\
$^{1}$Department of Theoretical Physics, Aristotle University of
Thessaloniki, \\ 54124 Thessaloniki, Greece  and \\
$^{2}$Theoretical Astrophysics, University of Tuebingen IAAT,\\
Auf der Morgenstelle 10, Tuebingen 72076 Germany}

\maketitle

\begin{abstract}
I study the effect of nuclear equation of state on the r-mode
instability of a rotating neutron star. I consider the case where
the crust of the neutron star is perfectly rigid and I employ the
related theory introduced by Lindblom {\it et al.}
\cite{Lidblom-2000}. The gravitational and the viscous time
scales, the critical angular velocity and the critical temperature
are evaluated by employing a phenomenological nuclear model for
the neutron star matter. The predicted equations of state for the
$\beta$-stable nuclear matter are parameterized by varying the
slope $L$ of the symmetry energy at saturation density on the
interval $72.5 \ {\rm MeV} \leq L \leq   110 \ {\rm MeV}$. The
effects of the density dependence of the nuclear symmetry energy
on r-mode instability  properties and the time evolution of the
angular velocity are presented and analyzed. A comparison of
theoretical predictions with observed neutron stars in low-mass
x-ray binaries (LMXBs) and millisecond radio pulsars (MSRPs) is
also performed and analyzed.  I estimate that it may be
possible to impose constraints on the nuclear equation of state, by a
suitable treatment  of observations and theoretical predictions of
the rotational frequency and spindown rate evolution  of known
neutron stars.

\vspace{0.3cm}

PACS number(s):
26.60.-c, 21.65.Ef, 04.40.Dg, 04.30.-w. \\

Keywords: Neutron stars; Nuclear equation of state; Nuclear
symmetry energy; R-mode instability; Gravitational waves.
\end{abstract}

%%%%%%%%%%%%%%%%%%%%%%%%%%%%%%%%%%%
\section{Introduction}
%%%%%%%%%%%%%%%%%%%%%%%%%%%%%%%%%%%%

The oscillations and instabilities of relativistic stars
\cite{Shapiro-83,Glendenning-2000,Haensel-07} gained a lot of
interest in the last decades because of the possible detection of
their gravitational waves
\cite{Andersson-1998,Friedman-98,Friedman-99,Andersson-2001,Andersson-2003,Kokkotas-99,Andersson-99,Kokkotas-2003,Andersson-00,Lindblom-2000-rev}.
Especially neutron stars may suffer a number of instabilities
which come in different flavors but they have a general feature in
common, they can be directly associated with unstable modes of
oscillation. In the present work I  concentrate my study on the so
called r-mode instability. The r-modes are oscillations of
rotating stars whose restoring force is the Coriolis force. The
gravitational radiation-driven instability of these modes has been
proposed as an explanation for the observed relatively low spin
frequencies of young neutron stars and of accreting neutron stars
in low-mass X-ray binaries as well \cite{Lidblom-2000}. This
instability can only occur when the gravitational-radiation
driving time scale of the r-mode is shorter than the time scales
of the various dissipation mechanisms that may occur in the
interior of the neutron star.

A very interesting problem is the consideration of the effect on
r-mode instability due to the presence of a solid crust in an old
neutron star. It is proved that the presence of a viscous boundary
layer under the solid crust of a neutron star increases the
viscous damping rate of the fluid r-modes
\cite{Lidblom-2000,Bildsten-2000}. Actually, the presence of  a
solid crust has a crucial effect on the r-mode motion and
following the discussion of Andersson and Kokkotas
\cite{Andersson-2001} this effect can be understood as follows:
based on the perfect fluid mode-calculations it is  anticipated the
transverse motion associated with the mode at the crust-core
boundary to be large. However, if the crust is assumed to be rigid,
the fluid motion must essentially fall off to zero at the base of
the crust in order to satisfy a non-slip condition (in the
rotating frame of reference).

Firstly, Bildsten and Ushomirsky~\cite{Bildsten-2000}, found that
the shear dissipation in the viscous boundary layer between the
solid crust and the fluid core decreases dramatically the viscous
damping time in cold old neutron stars as well as in hot young
neutron stars. They concluded that the r-mode instability is
unlikely to play a role in old, accreting neutron stars where in
hot young neutron stars the boundary-layer damping mechanism
limits the ability of the r-mode instability to reduce the angular
momentum of the star and consequently to produce detectable
amounts of gravitational radiation.

Anderson {\it et al.} \cite{Andersson-2000} used various
neutron-star parameters to obtain  significantly different results
for the critical frequency of the onset of r-mode instability.
They found  the critical velocity to be about 40$\%$ lower than
the estimate of Bildsten and Ushomirsky and inferred that the
r-mode instability is likely to be the mechanism that limits the
LMXBs spin periods and those of other millisecond pulsars as well
\cite{Andersson-2000}.

Rieutord \cite{Rieutord-2000} improved the model of the boundary
layer and found that the critical velocity agrees rather closely
with the original estimates of Bildsten and Ushomirsky. Lindblom
\cite{Lidblom-2000} improved previous estimates of the damping
rate by including the effect of the Coriolis force on the
boundary-layer eigenfunction, using more realistic neutron-star
models. They concluded that if the crust is assumed to be
perfectly rigid, the gravitational radiation driven instability in
the r-modes is completely suppressed in neutron stars colder than
about $1.5\times 10^8$ K and also found that the r-mode
instability is responsible for limiting the spin periods of the
LMXBs.

Wen {\it et al.}~\cite{Wen-012} studied the sensitivity of the
neutron star r-mode instability window to the density dependence
of the nuclear symmetry energy. Employing a simple model of a
neutron star with a perfectly rigid crust constructed with a set
of crust and core equations of state that span the range of
nuclear experimental uncertainty in the symmetry energy, they
concluded that smaller values of the slope parameter $L$ of the
symmetry energy help stabilize neutron stars against runaway
r-mode oscillations. Vidana~\cite{Vidana-012} analyzed also the
role of the symmetry energy slope parameter $L$ on the r-mode
instability by using both microscopic and phenomenological
approaches of the nuclear equation of state. He showed that the
r-mode instability region is smaller for those models which give
larger values of $L$. Alford~{\it et al.}~\cite{Alford-2012}
studied the viscous damping of r-modes of compact stars and
analyzed the regions where small amplitude modes are unstable to
the emission of gravitational radiation. They showed that many
aspects, such as the physically important minima of the
instability boundary, are surprisingly insensitive to detailed
microscopic properties of the considered form of matter.

Recently, Haskell  {\it et al.}~\cite{Haskell-012}
illustrated how current X-ray and ultra-violet observations can
constrain the physics of the r-mode instability. They discussed
various mechanisms active in the interior of a neutron
star and showed how these mechanisms can modify the instability
window in order to be consistent with the observations.

The motivation of the present work, in view of the above studies,
is to study extensively the nuclear equation of state (EOS) effect
on the r-mode instability and evolution of neutron stars with a
perfectly rigid crust
\cite{Lidblom-2000,Wen-012,Yoshida-2001,Levin-01}. Actually EOS
affects the time scales associated with the r-mode, in two
different ways. Firstly, EOS defines the radial dependence of the
mass density distribution $\rho(r)$, which is the basic ingredient
of the relevant integrals (see Sec.~II). Secondly, it defines the
core-crust transition density $\rho_c$ and also the core radius
$R_c$ which is the upper limit of the mentioned integrals. I
employ a phenomenological model for the energy per baryon  of the
asymmetric nuclear matter having the advantage of an analytical
form. By suitably choosing the parametrization of the model I
obtain various  forms for the density dependence of the nuclear
symmetry energy by varying the slope parameter $L$ in the interval
$65 \ {\rm MeV} \leq L \leq 110 \ {\rm MeV}$. The calculated EOS
concerns the neutron star core (from the center of the star up to
the crust-core interface). For the neutron star crust I employed
the EOS taken from the previous work of Feynman, Metropolis and
Teller \cite{Feynman-49} and also from Baym, Pethick and
Sutherland \cite{Baym-71}. The transition pressure at the edge of
the core is calculated by employing the thermodynamic method. It
is found that in a good approximation the quantities connected
with the r-mode instability are related directly with the slope
parameter. The effects of the stiffness of EOS are studied on the
gravitational  and viscous time scales as well as on the critical
angular velocity and critical temperature. I investigate the case
whether the observed properies of the LMXBs and MSRPs  should
constrain the slope parameter $L$.

The interesting issue is how much  the equation of state affects the
time evolution of the frequency and spindown rate of a  neutron star. This is also under consideration in the present work~\cite{Owen-98}.
In particular,  I develop my  study
on the evolution of r-mode by employing the mentioned EOS's and probing  the sensitivity  of the relevant quantities on the  slope parameter $L$.
A comparison of the theoretical predictions with a few observed cases is also presented and discussed.

It is worth  pointing out that in the present study I do not
include other additional dissipation mechanisms due for example to
bulk and  shear viscosity, which take into account the whole
star and not only the boundary layer at the crust core transition.
Actually, recently it was found that bulk and shear viscosities
increase with $L$ and therefore the damping of the mode is more
efficient for the models with  larger $L$~\cite{Vidana-012}. In
the present work I concentrate my study on the shear dissipation
in the viscous boundary layer between crust and core. The
mentioned dissipation mechanism decreases the viscous damping time
scale by more than a factor of $10^5$ in old, accreting neutron
stars and more than $10^7$ in hot young neutron stars and therefore becomes
comparable to the gravitational radiation driving time
scale~\cite{Lidblom-2000,Bildsten-2000}. These cases
efficiently limit  the ability of the r-mode instability to
reduce the angular momentum of the star and hence to produce
detectable gravitational radiation~\cite{Lidblom-2000}.

The article is organized as follows. In Sec~II I review briefly
the stability and evolution of the r-modes. Section III contains
the employed nuclear physics model  focusing on the presentation
of the nuclear symmetry energy and the thermodynamic method
applied for location of core-crust interface. The results are
presented and discussed in Sec. IV and Sec. V summarizes the
present study.

%%%%%%%%%%%%%%%%%%%%%%%%%%%%%%%%%%
\section{Stability and evolution of the r-modes}
%%%%%%%%%%%%%%%%%%%%%%%%%%%%%%%%%%%%

The {\it r}-modes evolve with a time dependence $ e^{i \omega
t-t/\tau}$ as a consequence of ordinary hydrodynamics and the
influence of the various dissipative processes. The real part of
the frequency of these modes, $\omega$, is given by
\cite{Lidblom-98}
\begin{equation}
\omega=-\frac{(m-1)(m+2)}{m+1}\Omega, \label{omega-1}
\end{equation}
where $\Omega$ is the angular velocity of the unperturbed star.
The imaginary part $1/\tau$ is determined by the effects of
gravitational radiation, viscosity, etc.
\cite{Lidblom-2000,Lidblom-98,Owen-98}. In the small-amplitude
limit, a mode is a driven, damped harmonic oscillator with an
exponential damping time scale
\begin{equation}
\frac{1}{\tau(\Omega,T)}=\frac{1}{\tau_{GR}(\Omega)}+\frac{1}{\tau_{bv}(\Omega,T)}+\frac{1}{\tau_{v}(\Omega,T)},
\label{t-1}
\end{equation}
where $\tau_{GR}$, $\tau_{bv}$, $\tau_{v}$ are gravitational
radiation, bulk viscosity and shear viscosity times scales respectively.
Gravitational radiation tends to drive the {\it r}-modes unstable,
while viscosity suppresses the instability. More precisely
dissipative effects cause the mode to decay exponentially as
$e^{-t/\tau}$ (i.e., the mode is stable) as long as $\tau >0$
\cite{Lidblom-98}.

The damping time $\tau_i$ for the individual mechanisms is defined
in general by \cite{Lidblom-2000,Vidana-012,Alford-2012}
\begin{equation}
\frac{1}{\tau_i}\equiv -\frac{1}{2E}\left(\frac{dE}{dt}\right)_i.
\label{tau-E-i}
\end{equation}
In Eq.~(\ref{tau-E-i})  the total energy $E$ of the r-mode is
given by
\begin{equation}
E=\frac{1}{2}\alpha^2R^{-2m+2}\Omega^2 \int_0^R
\rho(r)r^{2m+2} dr, \label{E-1}
\end{equation}
where $\alpha$ is the dimensionless amplitude of the mode, $R$ and
$\Omega$ are the radius and the angular velocity of the neutron
star respectively and $\rho(r)$ is the radial dependence of the
mass density of the neutron star. Similar expressions hold for
the dissipation rate $(dE/dt)_i$ \cite{Lidblom-98}.

The damping time scale due to viscous dissipation at the boundary
layer of the perfectly rigid crust and fluid core is given by
\cite{Lidblom-2000}
\begin{equation}
\tau_{v}=\frac{1}{2\Omega}\frac{2^{m+3/2}(m+1)!}{m(2m+1)!! {\cal
I}_m}\sqrt{\frac{2\Omega R_c^2\rho_c}{\eta_c}}
\int_0^{R_c}\frac{\rho(r)}{\rho_c}\left(\frac{r}{R_c}\right)^{2m+2}
\frac{dr}{R_c},
 \label{tv-1}
\end{equation}
where the quantities $R_c$, $\rho_c$ and $\eta_c$ are the radius,
density and viscosity of the fluid at the outer edge of the core respectively.
In deriving expression (\ref{tv-1}) it is assumed that the crust
is rigid and hence static in the rotating frame. The motion of the
crust due to the mechanical coupling to the core effectively
increases $\tau_v$ by a factor of $(\Delta v/v)^{-2}$, where
$\Delta v/v$ denotes the difference between the velocities in the
inner edge of the crust and outer edge of the core divided by the
velocity of the core \cite{Levin-01}.

In neutron stars colder than about $10^9$ K the shear viscosity is
expected to be dominated by electron-electron scattering. The
viscosity associated with this process is given by
\cite{Lidblom-2000}
\begin{equation}
\eta_{ee}=6.0\times 10^6 \rho^2 T^{-2}, \qquad ({\rm g \ cm^{-1}}
\ s^{-1}), \label{eta-ee-1}
\end{equation}
where all quantities are given in cgs units and $T$ is measured in
K. For temperature above $10^9$ K, neutron-neutron scattering
provides the dominant dissipation mechanism. In this range the
viscosity is given by \cite{Lidblom-2000}
\begin{equation}
\eta_{nn}=347 \rho^{9/4} T^{-2},\qquad ({\rm g\ cm^{-1}\ s^{-1}}).
\label{eta-nn-1}
\end{equation}
In the present work I neglect the effects of bulk viscosity,
which are not important for $T \leq 10^{10} \ K  $.

The fiducial viscous time scale $\tilde{\tau}_v$ is defined as
\begin{equation}
\tau_v=\tilde{\tau}_v\left(\frac{\Omega_0}{\Omega} \right)^{1/2}
\left(\frac{T}{10^8 \ K} \right), \label{fid-t-v}
\end{equation}
where $\Omega_0=\sqrt{\pi G\overline{\rho}}$ and
$\overline{\rho}=3M/4\pi R^3$ is the mean density of the star.

The gravitational radiation time scale is given by
\cite{Lidblom-2000,Lidblom-98}
\begin{equation}
\frac{1}{\tau_{GR}}=-\frac{32\pi G
\Omega^{2m+2}}{c^{2m+3}}\frac{(m-1)^{2m}}{[(2m+1)!!]^2}\left(\frac{m+2}{m+1}\right)^{2m+2}
\int_0^{R_c}\rho(r) r^{2m+2} dr, \label{tgr-1}
\end{equation}
%%%%%%
while the fiducial gravitational radiation time scale
$\tilde{\tau}_{GR}$ is defined as
\begin{equation}
\tau_{GR}=\tilde{\tau}_{GR}\left(\frac{\Omega_0}{\Omega}
\right)^{2m+2}.  \label{fid-t-GR}
\end{equation}
The critical angular velocity $\Omega_c$, above which the {\it
r}-mode is unstable, is defined by the condition
$\tau_{GR}=-\tau_{v}$ and is given, for $m=2$, by
\cite{Lidblom-2000,Lidblom-98}
\begin{equation}
\frac{\Omega_c}{\Omega_0}=\left(-\frac{\tilde{\tau}_{GR}}{\tilde{\tau}_v}
\right)^{2/11}\left(\frac{10^8 \ K}{T}  \right)^{2/11}.
\label{Omega-c-1}
\end{equation}
For a given temperature $T$ and mode $m$ the equation for the
critical angular velocity, that is $1/\tau(\Omega_c)=0$, is a
polynomial of order $m+1$ in $\Omega_c^2$, and thus each mode has
its own critical angular velocity \cite{Lidblom-98}. However, only
the smallest of these (the $m=2$ r-mode here) represents the
critical angular velocity of the star and I concentrate my study
on this r-mode.

Moreover, the maximum angular velocity $\Omega_K$ (Kepler angular
velocity) for any star occurs when the material at the surface
effectively orbits the star \cite{Lidblom-98}. This velocity is
nearly $\Omega_K= \frac{2}{3}\Omega_0$. Thus, there is a critical
temperature below which the gravitational radiation instability is
completely suppressed by viscosity and is given by
\cite{Lidblom-2000}
\begin{equation}
\frac{T_c}{10^8 \ K}=\left(\frac{\Omega_0}{\Omega_{c}}
\right)^{11/2}\left(-\frac{\tilde{\tau}_{GR}}{\tilde{\tau}_{v}}\right)=
\left(\frac{3}{2}\right)^{11/2}\left(-\frac{\tilde{\tau}_{GR}}{\tilde{\tau}_{v}}\right).
\label{Omega-c-2}
\end{equation}
Employing Eqs.~(\ref{Omega-c-1}) and (\ref{Omega-c-2}) the
critical angular velocity is expressed in terms of $T_c$, that is
\begin{equation}
\frac{\Omega_c}{\Omega_0}=\frac{\Omega_{K}}{\Omega_0}\left(\frac{T_c}{T}\right)^{2/11}=
\frac{2}{3}\left(\frac{T_c}{T}\right)^{2/11}. \label{Omega-c-3}
\end{equation}
Once the equation of state for the neutron star core and crust is
fixed, then all the ingredients of the r-mode instability, that is
the transition density $\rho_c$, the radial dependence of the mass
density $\rho(r)$, and the bulk properties of the neutron star
(mass, radius and core radius) are determined in a self-consistent
way.

Following the discussion of Owen {\it et al.}
\cite{Owen-98} during the phase where the angular momentum is radiated away to infinity by gravitational radiation, the angular velocity evolves as follows
\begin{equation}
\frac{d\Omega}{dt}=\frac{2\Omega}{\tau_{GR}}\frac{\alpha^2 Q}{1-\alpha^2Q},
\label{dOmegadt-1}
\end{equation}
where $\alpha$ is the dimensionless r-mode amplitude  parameter. In general, $\alpha$ which strongly affects the r-mode evolution is treated as a free parameter and usually varied on the large interval $\alpha=1-10^{-8}$. The quantity $Q$  related with the equation of state  is defined as $Q=3\tilde{J}/2\tilde{I}$ where
\begin{equation}
\tilde{J}=\frac{1}{MR^4}\int_{0}^R \rho(r)r^6 dr, \qquad \tilde{I}=\frac{8\pi}{2MR^2}\int_{0}^R \rho(r)r^4 dr.
\label{I-J-1}
\end{equation}
In order to complete the model for the evolution of the r-mode, one may take into account the thermal evolution, since temperature strongly influences
the dissipation mechanisms. The mentioned thermal evolution can be studied with an energy balance between the relevant radiative and viscous process, that is~\cite{Shapiro-83}
\begin{equation}
\frac{dT}{dt}=\frac{1}{C_v}\left(-L_{\nu}+H_s\right).
\label{ev-T-1}
\end{equation}
In Eq.~(\ref{ev-T-1}) $L_{\nu}$ is the neutrino luminosity, $H_s$ is the heating rate generated by shear viscosity and $C_v$ is the total heat capacity. In the present work,  I consider the approximate  case where the heating process balances the cooling one, that is $L_{\nu}=H_s$ (for a more detailed analysis see Ref.~\cite{Bondarescu-09}). In this
simplified case the solution of Eq.~(\ref{dOmegadt-1}) gives
\begin{equation}
\Omega(t)=\left(\frac{1}{\Omega^{-6}_{in} -{\cal C}t}\right)^{1/6},
\label{dOmegadt-2}
\end{equation}
where
\begin{equation}
{\cal C}=\frac{2\alpha^2Q}{\tilde{\tau}_{GR} (1-\alpha^2Q)}\frac{1}{\Omega_0^6}.
\label{Cons-1}
\end{equation}
$\Omega_{in}$ is a free parameter which corresponds to the initial angular velocity. The spindown rate of the angular velocity can be  easily derived from Eq.~(\ref{dOmegadt-2}) and is given by
\begin{equation}
\frac{d\Omega}{dt}=\frac{{\cal C}}{6}\left(\frac{1}{\Omega_{in}^{-6}-{\cal C}t}\right)^{7/6}.
\label{dOmegadt-3}
\end{equation}
%
%or after some algebra
%\begin{equation}
%\frac{d\Omega}{dt}=\frac{1}{6}\frac{{\cal C}\Omega_{in}^7}{(1-{\cal C}\Omega_{in}^6t)^{7/6}}
%\label{dOmegadt-4}
%\end{equation}
%
 A neutron star spin decreases  until it approaches its critical angular velocity $\Omega_c$. The time $t_c$ needed for $\Omega_{in}$ to evolve  to its minimum value $\Omega_c$
is
\begin{equation}
t_c=\frac{1}{{\cal C}}\left(\Omega_{in}^{-6}-\Omega_{c}^{-6}\right).
\label{tc-1}
\end{equation}

%%%%%%%%%%%%%%%%%%%%%%%%%%%%%%%%%%%
\section{The nuclear model}
%%%%%%%%%%%%%%%%%%%%%%%%%%%%%%%%%%%%%%%%%
The model used here, which has already been presented and analyzed
in previous papers
\cite{Prakash-97,Moustakidis-07-1,Moustakidis-07-2,Moustakidis-08,Moustakidis-09-1,Moustakidis-09-2},
is designed to reproduce the results of the microscopic
calculations of both nuclear and neutron-rich matter at zero
temperature and can be extended to finite temperature
\cite{Prakash-97,Moustakidis-07-2,Moustakidis-08,Moustakidis-09-1,Moustakidis-09-2}.
The energy per baryon  at $T=0$, is given by
\begin{eqnarray}
E_b(n,I)&=&\frac{3}{10}E_F^0u^{2/3}\left[(1+I)^{5/3}+(1-I)^{5/3}\right]+
\frac{1}{3}A\left[\frac{3}{2}-(\frac{1}{2}+x_0)I^2\right]u
\nonumber \\ &+&
\frac{\frac{2}{3}B\left[\frac{3}{2}-(\frac{1}{2}+x_3)I^2\right]u^{\sigma}}
{1+\frac{2}{3}B'\left[\frac{3}{2}-(\frac{1}{2}+x_3)I^2\right]u^{\sigma-1}}
 \label{e-T0}\\
&+&\frac{3}{2}\sum_{i=1,2}\left[C_i+\frac{C_i-8Z_i}{5}I\right]\left(\frac{\Lambda_i}{k_F^0}\right)^3
\left(\frac{\left((1+I)u\right)^{1/3}}{\frac{\Lambda_i}{k_F^0}}-
\tan^{-1} \frac{\left((1+
I)u\right)^{1/3}}{\frac{\Lambda_i}{k_F^0}}\right)\nonumber \\
&+&
\frac{3}{2}\sum_{i=1,2}\left[C_i-\frac{C_i-8Z_i}{5}I\right]\left(\frac{\Lambda_i}{k_F^0}\right)^3
\left(\frac{\left((1-I)u\right)^{1/3}}{\frac{\Lambda_i}{k_F^0}}-
\tan^{-1}
\frac{\left((1-I)u\right)^{1/3}}{\frac{\Lambda_i}{k_F^0}}\right)
\nonumber.
%\label{Eb-eos}
\end{eqnarray}
In Eq.~(\ref{e-T0}), $I$ is the asymmetry parameter
($I=(n_n-n_p)/n$) and $u=n/n_0$, with $n_0$ denoting the
equilibrium symmetric nuclear matter density, $n_0=0.16$
fm$^{-3}$.   The parameters $A$, $B$, $\sigma$, $C_1$, $C_2$ and
$B'$ which appear in the description of symmetric nuclear matter
are determined in order that $E_b(n=n_0,I=0)=-16$ {\rm MeV},
$n_0=0.16$ fm$^{-3}$, and the incompressibility is $K=240$ {\rm
MeV} and have the values $A=-46.65$, $B=39.45$, $\sigma=1.663$,
$C_1=-83.84$, $C_2=23$ and $B'=0.3$. The finite range parameters
are $\Lambda_1=1.5 k_F^{0}$ and $\Lambda_2=3 k_F^{0}$ and $k_F^0$
is the Fermi momentum at the saturation point $n_0$. The baryon
energy is written also as a function of the baryon density $n$ and
the proton fraction $x$ ($x=n_p/n$), that is $E_b(n,x)$, by
replacing $I=1-2x$.

%%%%%%%%%%%%%%%%%%%%%%%%%%%%%%%%%%%%%%%%%%%%%%%%%%%%%%%%%%%%%%%%%%%%
The additional parameters $x_0$, $x_3$, $Z_1$, and $Z_2$ employed
to determine the properties of asymmetric nuclear matter are
treated as parameters constrained by empirical knowledge
\cite{Prakash-97}. The parameterizations used in the present model
have only a modest microscopic foundation. Nonetheless, they have
the merit of being able to closely approximate more physically
motivated calculations as presented in Fig.~1. More precisely, in
Fig.~1 I compare the energy per baryon (for symmetric nuclear
matter (Fig.~1a) and pure neutron matter (Fig.~1b)) calculated by
the present schematic model referred to  as momentum-dependent
interaction model (MDIM), with those of exististing, state of the art
calculations by Wiringa et al. \cite{Wiringa-88-b} and
Pandharipande et al. \cite{Pandharipande-98}.

%%%%%%%%%%%%%%%%%%%%%%%%%%%%%%%%%
\subsection{Symmetry energy}
%%%%%%%%%%%%%%%%%%%%%%%%%%%%%%%%
The energy $E_b(n,I)$ can be expanded around $I=0$ as follows
\begin{equation}
E_b(n,I)=E_b(n,I=0)+E_{sym,2}(n)I^2+E_{sym,4}(n)I^4+\dots
+E_{sym,2k}(n)I^{2k}+\dots, \label{Expans-1}
\end{equation}
where  the coefficients of the expansion are given by the
expression
\begin{equation}
E_{sym,2k}(n)=\left.
\frac{1}{(2k)!}\frac{\partial^{2k}E_b(n,I)}{\partial
I^{2k}}\right|_{I=0}. \label{Expan-2}
\end{equation}
In (\ref{Expans-1}), only even powers of $I$ appear due to the
fact that the strong interaction must be symmetric under exchange
of  neutrons with protons i.e.  the contribution to the energy
must be independent of the sign of the difference $n_n-n_p$. The
nuclear symmetry energy $E_{sym}(n)$ is defined as the coefficient
of the quadratic term, that is
\begin{equation}
E_{sym}(n)=E_{sym,2}(n)=\left.
\frac{1}{2!}\frac{\partial^{2}E_b(n,I)}{\partial
I^{2}}\right|_{I=0}. \label{esym-1}
\end{equation}
The slope of the symmetry energy $L$  at nuclear saturation
density $n_0$, which  is correlated with the crust-core transition
density $n_t$ in a neutron star, is defined as
\begin{equation}
L=3n_0 \left. \frac{d E_{sym}(n)}{d
n}\right|_{n=n_0}. \label{L-1}
\end{equation}
By suitably choosing the parameters $x_0$, $x_3$, $Z_1$, and
$Z_2$, it is possible to obtain different forms for the density
dependence of the symmetry energy $E_{sym}(n)$ as well as on the
value of the slope parameter $L$. I take as a range of $L$ $65 \
{\rm MeV}  \leq L \leq 110 \ {\rm MeV} $ where the value of the
symmetry energy at saturation density is fixed to be
$E_{sym}(n_0)=30$ MeV. Actually, for each value of $L$ the density
dependence  of the symmetry energy is adjusted so that the energy
of pure neutron matter is comparable with those of existing
state-of-the-art calculations
\cite{Wiringa-88-b,Pandharipande-98}.

Fig.~2(a) displays the behavior of the nuclear symmetry energy as
a function of the ratio $u=n/n_0$ for various values of the slope
parameter $L$. The aim of the above simple parametrization is to
reproduce the nuclear symmetry energy originating from more
realistic microscopic calculations and also covers the possible
range of the nuclear symmetry energy dependence on the density.

%%%%%%%%%%%%%%%%%%%%%%%%%%%%%%%%%
\subsection{Proton fraction}
%%%%%%%%%%%%%%%%%%%%%%%%%%%%%%%%

I examine the  proton fraction $x$ (as a function of the baryon density $n$)
in $\beta$-stable matter. In this case the following processes take place simultaneously
\begin{equation}
n \longrightarrow p+e^{-}+\bar{\nu}_e \qquad \qquad p +e^{-}
\longrightarrow n+ \nu_e,
\end{equation}
I assume that neutrinos
generated in these reactions have left the system. This implies
that
\begin{equation}
\hat{\mu}=\mu_n-\mu_p=\mu_e. \label{chem-1}
\end{equation}
The demand for equilibrium leads to equation
\begin{equation}
\frac{\partial }{\partial x}\left(\frac{}{}
E_b(n,x)+E_e(n,x)\right)=0, \label{equil-1}
\end{equation}
or
\begin{equation}
\left( \frac{\partial E_b}{\partial x}\right)_n=-\left(
\frac{\partial E_e}{\partial x}\right)_n=-\mu_e. \label{x-frac-1}
\end{equation}
Finally, considering that the chemical potential of the electron
is given by the relation (relativistic electrons)
\begin{equation}\mu_e=\sqrt{k_{F_e}^2c^2+m_e^2c^4}\simeq k_{F_e}
c=\hbar c(3 \pi^2 n x)^{1/3}, \label{chem-ele-1}
\end{equation}
then Eq.~(\ref{x-frac-1}) is written
\begin{equation} \left(
\frac{\partial E_b}{\partial x}\right)_n=-\hbar c(3 \pi^2 n
x)^{1/3}. \label{x-frac-2}
\end{equation}
Eq.~(\ref{x-frac-2}) determines the proton fraction of
$\beta$-stable matter. In Fig.~2(b) I plot the proton fraction
calculated from equation (\ref{x-frac-2}) as a function of the
ratio $u=n/n_0$ for various values of the slope parameter $L$.
According to Fig.~2(b) for low values of $L$, the neutron star
matter consists mainly from neutrons and only a very small
fraction of protons. However the increase of $L$ (and consequently
the increase of the stiffness of EOS) leads to an increase of the
proton fraction. The values of the proton fraction define the kind
of the cooling process of a hot neutron star. More precisely, it
is well known that the direct Urca process can occur in neutron
stars if the proton concentration exceeds the critical value
$x_{crit}=0.11$ for neutron star matter with electrons and
$x_{crit}=0.148$ when electrons and muons coexist in neutron star
matter~ \cite{Prakash-94}.

%%%%%%%%%%%%%%%%%%%%%%%%%%%%%%%%%%%%%%%%%%%%%%%%%%%%%%%%%%%%%%%%%%
\subsection{Nuclear equation of state for $\beta$-stable matter}
%%%%%%%%%%%%%%%%%%%%%%%%%%%%%%%%%%%%%%%%%%%%%%%%%%%%%%%%%%%%%%%%%
The total pressure $P(n,x)$, in the core of a neutron star, is
decomposed into baryon and electron contributions
\begin{equation}
P(n,x)=P_b(n,x)+P_e(n,x), \label{P-all-1}
\end{equation}
where
\begin{equation}
P_b(n,x)=n^2\frac{\partial E_b(n,x)}{\partial  n}. \label{Pb-1}
\end{equation}
The electrons are considered as a non-interacting relativistic
Fermi gas and their contribution to the total energy density
$\epsilon_{e}(n,x)$ and pressure $P_e(n,x)$  reads
\begin{equation}
\epsilon_{e}(n,x)=\frac{\hbar c}{4 \pi^2}\left(3\pi^2
xn\right)^{4/3}, \label{ele-ene}
\end{equation}
\begin{equation}
P_e(n,x)=\frac{\hbar c}{12 \pi^2}\left(3\pi^2 xn\right)^{4/3}.
\label{Pe-2}
\end{equation}
Now the total energy density $\epsilon_{tot}$ and pressure
$P_{tot}$ of charge neutral and chemically equilibrium nuclear
matter is
\begin{equation}
\epsilon_{tot}=\epsilon_b+\epsilon_e, \label{tot-e}
\end{equation}
\begin{equation}
P_{tot}=P_b+P_e \ .\label{tot-pr}
\end{equation}
From Eqs.~(\ref{tot-e}) and  (\ref{tot-pr}) I construct the
equation of state in the form $\epsilon=\epsilon(P)$. When the
electrons energy is large enough (i.e. greater than the muon
mass), it is energetically favorable for the electrons to convert
to muons
\begin{equation}
e^{-} \longrightarrow \mu^{-}+\bar{\nu}_{\mu}+\nu_e.
\label{tot-endens}
\end{equation}
However, in the present work I will not include the muon case to
the total equation of state since the muon contribution does not
alter significantly the gross properties of the neutron stars.

%%%%%%%%%%%%%%%%%%%%%%%%%%%%%%%%%%%%%%%
\subsection{The thermodynamical method}
%%%%%%%%%%%%%%%%%%%%%%%%%%%%%%%%%%%%%%%%%
The core-crust interface corresponds to the phase transition
between nuclei and uniform nuclear matter. The uniform matter is
nearly pure neutron matter, with a proton fraction of just a few
percent determined by the condition of beta equilibrium. Weak
interactions conserve both baryon number and charge
\cite{Lattimer-07}, and from the first law of thermodynamics, at
temperature $T=0$ I have
\begin{equation}
{\rm d}{\cal E}=-P{\rm d}v-\hat{\mu}{\rm d}q, \label{u-1}
\end{equation}
where ${\cal E}$ is the internal energy per baryon, $P$ is the total
pressure,  $v$ is the volume per baryon ($v=1/n$ where $n$ is the
baryon density) and $q$ is the charge fraction ($q=x-Y_e$ where
$x$ and $Y_e$ are the proton and electron  fractions in baryonic
matter respectively). In $\beta$-equilibrium the chemical
potential $\hat{\mu}$ is given by $\hat{\mu}=\mu_n-\mu_p=\mu_e$
where $\mu_p$, $\mu_n$ and $\mu_e$  are the chemical potentials of
the protons, neutrons and electrons respectively.  The stability
of the uniform phase requires that ${\cal E}(v,q)$ is a convex
function~\cite{Callen-85}. This condition leads to the following
two constraints for the pressure and the chemical potential
\begin{equation}
-\left(\frac{\partial P}{\partial v}\right)_q-\left(\frac{\partial
P}{\partial q}\right)_v \left(\frac{\partial q}{\partial
v}\right)_{\hat{\mu}}>0, \label{cond-1}
\end{equation}
\begin{equation}
-\left(\frac{\partial \hat{\mu}}{\partial q}\right)_v>0.
\label{cond-2}
\end{equation}
It is assumed that the total internal energy per baryon ${\cal E}(v,q)$
can be decomposed into baryon ($E_N$) and electron ($E_e$)
contributions
\begin{equation}
{\cal E}(v,q)=E_b(v,q)+E_e(v,q). \label{u-2}
\end{equation}
The relative theory has been extensively presented in the recent
publication~\cite{Moustakidis-010}. I consider the condition of
charge neutrality $q=0$ which requires that $x=Y_e$. This is the
case  which will also be taken into account  in the present study. Hence, according
to Ref.~\cite{Moustakidis-010} the constraints (\ref{cond-1}) and
(\ref{cond-2}), after some algebra lead to the following
constraint
\begin{equation}
C(n)=2n\frac{\partial E_b(n,x)}{\partial n}+n^2\frac{\partial
^2E_b(n,x)}{\partial n^2}-\left(\frac{\partial^2E_b(n,x)}{\partial
n
\partial x}n \right)^2\left(\frac{\partial^2E_b(n,x)}{\partial x^2}
\right)^{-1} > 0,  \label{cont-1}
\end{equation}
For a given equation of state, the quantity $C(n)$ is plotted as a
function of the baryon density $n$ and the equation $C(n)=0$
defines the transition density $n_t$.

%%%%%%%%%%%%%%%%%%%%%%%%%%%%%%%%
\section{Results and Discussion}
%%%%%%%%%%%%%%%%%%%%%%%%%%%%%%%%%
I employ a phenomenological model for  the energy per baryon  of
the asymmetric nuclear matter having the advantage of an
analytical form. By suitably choosing the parametrization of the
model I  obtain various forms for the density dependence of the
energy per baryon of neutron matter (see Fig.~1(b)), the nuclear
symmetry energy (see Fig.~2(a)) and the proton fraction (see
Fig.~2(b)), and in total  the neutron star core equation of state.

In order to clarify further the effect of the symmetry energy on the proton fraction I plot in  Fig.~2(c) the density dependence of the fourth  order term $E_{sym,4}(n)$ (see the expansion (\ref{Expans-1})) for the various values of $L$. Considering, for example a fourth order approximation
on the symmetry energy  and combining Eqs.~(\ref{x-frac-2}) and  (\ref{Expans-1}) I found that the density dependence of the proton fraction is determined by the equation
\[4(1-2x)E_{sym,2}(n)+8(1-2x)^3E_{sym,4}(n)\simeq \hbar c(3\pi^2nx)^{1/3}. \]
Obviously, the second order term $E_{sym,2}(n)$ mostly affects the proton fraction density dependence. However, the contribution of the fourth order term $E_{sym,4}(n)$ is not negligible and in some cases (see for example the case $L=65$) has a significant effect on $x(n)$.

Additionally, the present model is applied for the determination
of the transition density $n_t$ between the crust and the core. In
order to complete the equation of state which describes the
neutron star matter for densities lower than the transition
density $n_t$ (equation of state of the crust) I employed the
relative equation of state of  Feynman, Metropolis and Teller
\cite{Feynman-49} and also of Baym, Pethick and Sutherland
\cite{Baym-71}.

In Fig.~3(a), I plot the transition density $n_t$, which is a
fundamental quantity in the present study, as a function of the
slope parameter $L$. A linear relation is found of the form
\begin{equation}
n_t=0.1253-0.0007\ L \ ({\rm fm^{-3}}),  \label{nt-L-1}
\end{equation}
where the slope parameter $L$ is given in MeV. According to relation (\ref{nt-L-1}) an increase of $L$ ( a stiffer
equation of state) leads to a decrease of $n_t$, to the extension of both the total radius $R$ (since the mass is fixed) and the  radius core $R_c$.  In Fig.~3(b), I plot also the dependence of the transition pressure
$P_t$ on $L$. Actually, during the last years there is an extensive
interest for the study of the transition density and pressure
in neutron stars~\cite{Moustakidis-010,Xu-09-2,Ducoin-010,Ducoin-011}. It was found that the transition density and pressure decrease roughly
linearly with the slope parameter $L$ using dynamical and thermodynamical methods~\cite{Moustakidis-010,Xu-09-2}. The authors in Refs.~\cite{Ducoin-010,Ducoin-011} used a large number of nuclear models and evaluated the dispersion affecting the correlation between the transition pressure $P_t$ and $L$. From a detailed analysis it was shown that this correlation is weak but $P_t$ is mainly correlated with the symmetry energy slope $L$ and curvature $K_{sym}$ defined at $\rho=0.1$ fm$^{-3}$~\cite{Ducoin-010,Ducoin-011}.

The transition pressure does not affect directly the
time scales corresponding to the r-mode instability. However it
influences significantly the crustal fraction of the  moment of
inertia $\Delta I/I$ \cite{Link-99}. The ratio $\Delta I/I$ is
particularly interesting as it can be inferred from observations
of pulsar glitches that is occasional disruptions of the otherwise
extremely regular pulsations from magnetized, rotating neutron
stars~\cite{Xu-09-2}.

In Fig.~4 I display the energy density $E_{ns}$ of neutron star matter, including both the fluid core and the solid crust matter,  as a function of the pressure $P$ for various values of the
slope parameter $L$. The mentioned equations of state are employed in order to solve the Tolman-Oppenheimer-Volkoff equations to calculate the bulk neutron star properties that is the mass, radius, density distribution of the baryonic matter e.t.c. which are the basic "ingredients" for the study of the r-mode instability and evolution.

In Fig.~5 I display the mass-radius relation for neutron stars
for the selected equation of states. All of them predict maximum
mass for neutron stars even higher than $1.8$M$_{\odot}$. To further
illustrate the mass-radius relation  I present in Table I and II
the transition density $n_t$, transition pressure $P_t$, the total
radius of the star $R$, the core radius $R_c$, the core mass $M_c$
as well as the central pressure $P_c$ for various values of the
parameter $L$ for neutron star mass $1.4$M$_{\odot}$ and
$1.8$M$_{\odot}$ respectively .

Considering that $\epsilon (r)$ is the energy
density function defined as $\epsilon(r)=\rho(r)/c^2$, then the integral $\displaystyle \int_{0}^{R_c} \rho (r) r^6
dr$,  which  is  a basic ingredient of the r-mode energy $E$ given
by Eq.~(\ref{E-1}) and also of the time scales (Eqs.~(\ref{tv-1})
and (\ref{tgr-1}) respectively),  can be written in the dimensionless form
\[I(R_c)=\int_{0}^{R_c}\left(\frac{\epsilon
(r)}{{\rm MeV \ fm^{-3} }}\right) \left(\frac{r}{{\rm km}}\right)^6 d\left(\frac{r}{{\rm km}}\right).   \]
The integral $I(R_c)$ is plotted in Fig.~6, as a
function of the slope parameter $L$ for the interval $72.5\ {\rm
MeV} \leq L \leq 110 \ {\rm MeV}$.  It is
obvious that the values of $I(R_c)$ are correlated almost linearly
with $L$ in the considered interval. The least-squares fit
expressions are respectively
\begin{equation}
I(R_c)=\left(\frac{}{}0.7+0.084 L\right)10^8, \quad (M=1.4
M_{\odot}) \label{Int-1}
\end{equation}
\begin{equation}
I(R_c)=\left(\frac{}{}1.24+0.0845 L\right)10^8, \quad (M=1.8
M_{\odot}) \label{Int-2}
\end{equation}
where the slope parameter $L$ is given in MeV.

The fiducial gravitational radiation time scale
$\tilde{\tau}_{GR}$  combining Eqs.~(\ref{tgr-1}) and
(\ref{fid-t-GR}) and after some algebra takes the form
\begin{equation}
\tilde{\tau}_{GR}=-0.7429 \left(\frac{R}{ {\rm km}}\right)^9
\left(\frac{1 M_{\odot}}{M}\right)^3 \left[I(R_c)\right]^{-1} \ ({\rm sec}), \label{taugr-1}
\end{equation}
where $R$,$r$  are given in km  and  $M$ in $ M_{\odot}$. I can
proceed further by putting Eq.~(\ref{Int-1}) into (\ref{taugr-1}).
In this way the time scale $\tilde{\tau}_{GR}$ corresponds to a
neutron star with mass $1.4$ M$_{\odot}$  given by the simple
expression
\begin{equation}
\tilde{\tau}_{GR}=-2.70736 \left(\frac{R}{10 {\rm
km}}\right)^9\frac{1}{0.7+0.084 L} \ ({\rm sec}). \label{taugr-1-14M}
\end{equation}
It is worth to point out that $\tilde{\tau}_{GR}$, according to
Eq.~(\ref{taugr-1-14M}) depends directly on the slope parameter
$L$, as well as indirectly on the values of the radius $R$. In any
case, Eq.~(\ref{taugr-1-14M}) exhibits the equation of state
dependence of the time scale $\tilde{\tau}_{GR}$ in a quantitative
way.

As a first approximation I can consider the case where the mass
density of the neutron star $\rho(r)$ is uniform that is
$\rho(r)=\overline{\rho}\equiv 3M/4\pi R^3$. Actually, this case
is unphysical because the energy density does not vanish on the
surface and the speed of sound $c_s=\sqrt{\partial P/\partial
\rho}$ is infinite \cite{Lattimer-01}. However, I consider this
case since it has been  extensively applied for r-mode
instabilities studies. After some algebra
$\tilde{\tau}_{GR}^{approx}$ is written
\begin{equation} \tilde{\tau}_{GR}^{approx}=-1.95
\left(\frac{R}{10 {\rm km}}\right)^{12}\left(\frac{10 {\rm
km}}{R_c} \right)^7\left( \frac{1 M_{\odot}}{M}\right)^4 \ ({\rm sec}).
\label{taugr-2}
\end{equation}
In Fig.~7, as well as in Table III,  I compare the values of
$\tilde{\tau}_{GR}$ and $\tilde{\tau}_{GR}^{approx}$. Actually
both are increasing functions of the slope parameter $L$, for low
values of $L$, but for higher values show a saturation trend. The
approximated values for gravitational time scales are lower
between $8\%-16\%$ (for $M=1.4 M_{\odot}$) and lower by
$23\%-28\%$ (for $M=1.8 M_{\odot}$). Consequently, the mean
density approximation, concerning the gravitational time scales
works better for low neutron star mass values.

The fiducial viscus time $\tilde{\tau}_{v}$ which results from
Eqs.~(\ref{tv-1}), (\ref{eta-ee-1}), (\ref{eta-nn-1}) and
(\ref{fid-t-v})  after some algebra is written for the case of
viscosity due to  electron-electron  and neutron-neutron
scattering respectively
\begin{equation}
\tilde{\tau}_{ee}=0.1446\cdot 10^8 \left(\frac{R}{ {\rm
km}}\right)^{3/4}\left(\frac{1 M_{\odot}}{M} \right)^{1/4}
\left(\frac{ {\rm km}}{R_c} \right)^6
\left(\frac{{\rm gr \ cm^{-3}}}{\rho_c}\right)^{1/2}\left(\frac{{\rm MeV \ fm^{-3} }}{\epsilon_c}\right) \ I(R_c) \ ({\rm sec}), \label{t-ee-2}
\end{equation}
\begin{equation}
\tilde{\tau}_{nn}=19\cdot 10^8
 \left(\frac{R}{
{\rm km}}\right)^{3/4}\left(\frac{1 M_{\odot}}{M} \right)^{1/4}
\left(\frac{ {\rm km}}{R_c} \right)^6
\left(\frac{{\rm gr \ cm^{-3}}}{\rho_c}\right)^{5/8}\left(\frac{{\rm MeV \ fm^{-3} }}{\epsilon_c}\right) \ I(R_c) \ ({\rm sec}) \label{t-nn-2}
\end{equation}
where $\rho_c$ and $\epsilon_c$ are the mass density (in gr
cm$^{-3}$) and the energy density (in MeV fm$^{-3}$) respectively
at the core edge. It is found that an almost linear relation holds between
the mass transition density $\rho_c$ and the slope parameter $L$
(and consequently a similar relation between $\epsilon_c$ and $L$), for the interval $72.5 \ {\rm MeV} \leq L \leq
110 \ {\rm MeV}$, with the form
\begin{equation}
\rho_c=\left(\frac{}{} 2.148-0.0125 L \right)\times 10^{14} \ \left({\rm
gr}\cdot {\rm cm}^{-3} \right), \label{rhoc-L}
\end{equation}
\begin{equation}
\epsilon_c=\left(\frac{}{} 2.148-0.0125 L \right)\times 56.1837 \
\left( {\rm MeV}\cdot {\rm fm}^{-3}\right). \label{epsilonc-L}
\end{equation}
Combining Eqs.~(\ref{t-ee-2}), (\ref{t-nn-2})  (\ref{rhoc-L}) and
(\ref{epsilonc-L}) I found for a neutron star mass with $M=1.4
M_{\odot}$ and  for the interval  ($72.5 \ {\rm MeV} \leq L \leq
110 \ {\rm MeV}$), the simple expressions
\begin{equation}
\tilde{\tau}_{ee}=13.3053\left(\frac{R}{10 {\rm km}}\right)^{3/4}
\left(\frac{10 {\rm km}}{R_c} \right)^6\frac{0.7+0.084
L}{(2.148-0.0125L)^{3/2}} \ ({\rm sec}) \quad (M=1.4 M_{\odot})
 \label{t-ee-2-L}
\end{equation}
\begin{equation}
\tilde{\tau}_{nn}=31.09\left(\frac{R}{10 {\rm km}}\right)^{3/4}
\left(\frac{10 {\rm km}}{R_c} \right)^6\frac{0.7+0.084
L}{(2.148-0.0125L)^{13/8}} \ ({\rm sec})  \quad (M=1.4 M_{\odot}).
 \label{t-nn-2-L}
\end{equation}
By employing the mean density approximation the corresponding
expressions are
\begin{equation}
\tilde{\tau}_{ee}^{approx}=55.1657\left(\frac{10 {\rm km}}{R}
\right)^{9/4}\left(\frac{R_c}{10 {\rm km}} \right)
\left(\frac{M}{1 M_{\odot}} \right)^{3/4}\left(\frac{{\rm gr}\cdot
{\rm cm}^{-3}}{\rho_{c,14}}\right)^{3/2}  \ ({\rm sec}) \label{t-ee-3}
\end{equation}
\begin{equation}
\tilde{\tau}_{nn}^{approx}=129\left(\frac{10 {\rm km}}{R}
\right)^{9/4}\left(\frac{R_c}{10 {\rm km}} \right)
\left(\frac{M}{1 M_{\odot}} \right)^{3/4}\left(\frac{{\rm gr}\cdot
{\rm cm}^{-3}}{\rho_{c,14}}\right)^{13/8}  \ ({\rm sec}), \label{t-nn-3}
\end{equation}
where $\rho_{c,14}$ is given in $10^{14}{\rm gr}\cdot {\rm
cm}^{-3}$. In Fig.~7, as well as in Table III,  I compare the
fiducial viscous time scales. Both time scales are an increasing
function of the slope parameter $L$, that is a stiffer equation of state leads to  smaller viscosity effects on the r-mode. The
approximated values for the viscosity timescales  are between
$9\%-16\%$ (for $M=1.4 M_{\odot}$) and between $23\%-28\%$ (for
$M=1.8 M_{\odot}$). Consequently, the mean density approximation,
concerning the viscosity time scales, works better for low values
of neutron stars.

In Fig.~8 are displayed the r-mode instability windows (which  specified  by the $\Omega_c/\Omega_0$-$T$ dependence)   for neutron
stars with mass $1.4 M_{\odot}$ and $1.8 M_{\odot}$ for the
selected equations of state as a function of the temperature. For
low values of temperature ($T \leq 10^9$ K) I plot the ratio
\begin{equation}
\frac{\Omega_c}{\Omega_0}=\left(-\frac{\tilde{\tau}_{GR}}{\tilde{\tau}_{ee}}
\right)^{2/11}\left(\frac{10^8 \ K}{T}  \right)^{2/11},
\label{Omega-c-ee}
\end{equation}
while for $T \geq 10^9$ K I plot the ratio
\begin{equation}
\frac{\Omega_c}{\Omega_0}=\left(-\frac{\tilde{\tau}_{GR}}{\tilde{\tau}_{nn}}
\right)^{2/11}\left(\frac{10^8 \ K}{T}  \right)^{2/11}.
\label{Omega-c-nn}
\end{equation}
The most striking feature is the location of the ratio
$\Omega_c/\Omega_0$ in a narrow interval (mainly in the case of
neutron star with mass $1.4 M_{\odot}$). Actually, the ratio
$\Omega_c/\Omega_0$ increases around $7.7\%$ (for $T \leq 10^9$ K)
and around $8.7\%$ (for $T \geq 10^9$ K) with the lower values
corresponding to the case of $L=110$ MeV and the higher to the case
$L=80$ MeV. It is concluded that the values of that ratio saturate
for $L$ close to the value $80$ MeV. In the case of a neutron star
with mass $M=1.8 M_{\odot}$ the ratio $\Omega_c/\Omega_0$ increases
around $5.5\%$ (for $T \leq 10^9$ K) and around $6.5\%$ (for $T \geq 10^9$ K)
with the lower values corresponding to the case of $L=110$
MeV and the higher to the case $L=80$ MeV.

Moreover I study  the effect of the nuclear equation of state
on  $\Omega_c$. In particular, I examine the sensitivity of
$\Omega_c$ on the density dependence of symmetry energy. Thus,  by
combining Eqs.~(\ref{taugr-1}), (\ref{t-ee-2}) and
(\ref{Omega-c-ee}), for a fixed neutron star mass and temperature,
I found the relation
\begin{equation}
\Omega_c \sim \frac{R_c^{12/11}}{(I(R_c))^{4/11}}\rho_c^{3/11}.
\label{Omegac-approx}
\end{equation}
It is obvious, from Eq.~(\ref{Omegac-approx}), that $\Omega_c$ is
sensitive to the structure (due to the factor
$I(R_c)$), size (due to $R_c$) and the interface edge (due to
$\rho_c$)  of the  core of the neutron star. Consequently,
$\Omega_c$  is sensitive  both to the high density dependence of
the EOS via the factors $R_c$  and $I(R_c)$   as well as to the
low density dependence of the symmetry energy via the factor
$\rho_c$. In addition, and summing up it is  shown that $\Omega_c$
is inversely  proportional to the core radius $R_c$ (the higher
value of $L$ the lower is the  value of $\Omega_c$) and proportional to
$\rho_c^{3/11}$ (a higher values of $L$ corresponds  to a lower  values of
$\rho_c$). It is worthwhile to notice here  that the dependence  of
$\rho_c$  on  the slope parameter $L$ is model dependent. It has been
found recently that the error due to the assumption that a priori
the equation of state is parabolic may introduce a large error in
the determination of related properties of neutron stars as the
crustal fraction of the moment of inertia and the critical
frequency of rotating neutron stars \cite{Xu-09-2,Moustakidis-12}.

The critical temperature $T_c$, defined by Eq.~(\ref{Omega-c-2}),
is plotted in Fig.~9, as a function of the slope parameter $L$  for neutron star with mass
$1.4 M_{\odot}$ and 1.8 $M_{\odot}$. $T_c$ is in the range from  $0.73\times 10^8$ to $1.1\times 10^8$ K (for 1.4 $M_{\odot}$ neutron star mass) and from  $0.18\times 10^8$ to $0.24 \times 10^8$ (for 1.8 $M_{\odot}$ neutron star mass) where  the maximum values  of $T_c$ correspond to equation of state  with $L=80$ MeV. In particular, $T_c$ is a decreasing function of $L$ (for $L  \geq 80$ MeV). The present finding is in contradiction  with the corresponding one presented in Ref.~\cite{Wen-012} where $T_c$ is a monotonously increasing function of $L$  for a large interval ($25 \leq L \leq 105$ MeV). The above disagreement  may be attributed with the use of two different models. More precisely, $T_c$, according to Eq.~(\ref{Omega-c-2}), depends on the interplay between $\tilde{\tau}_{GR}$ and   $\tilde{\tau}_{ee}$, and consequently, for fixed neutron star mass, on the parameter $L$. Obviously, this interplay is model dependent and related with the $L$ dependence on the  fiducial times scales as exhibited in Fig.~7 of the present work and  in Fig.~3 of Ref.~\cite{Wen-012}. Although, in contrast to $T_c$-$L$ dependence the present results are quantitatively in  very good  agrement with those in Ref.~\cite{Wen-012}, at least for the 1.4 $M_{\odot}$ neutron star mass and for the same range of $L$.

In Fig.~10 I compare the r-mode instability window for the
selected equations of state with those of the observed neutron
stars in LMXBs and MSRPs for $M=1.4 M_{\odot}$ and $M=1.8
M_{\odot}$. I find that the instability window drops by $\simeq
20-40 $ Hz when the mass is raised from $M=1.4 M_{\odot}$ to
$M=1.8 M_{\odot}$. In addition, the stiffness equation of state leads an  increase of  the instability window (which specified, in this case,  by the $\nu_c$-$T$ dependence).
Following the study of Wen {\it et al.}~\cite{Wen-012} and Haskell
{\it et al.}~\cite{Haskell-012} I include many cases of LMXBs and
a few of MSRPs (for more details see \cite{Watts-08,Keek-010} and
Table~1 of Ref.~\cite{Haskell-012}). The masses of the mentioned
stars are not measured accurately. In addition, it is worth pointing out that the estimates of the core temperature have large uncertainties. In the present work, the core temperatures $T$ are taken from Ref.~\cite{Haskell-012} and the uncertainties, in a few relevant cases, are derived by employing  the method suggested in Ref.~\cite{Ho-011}.  In particular, the core temperatures $T$  are derived by combining their observed accretion luminosity and considering that the cooling is dominated by the modified Urca neutrino process for normal nucleons (lower limit of $T$) or by the modified Urca, taking into account the effect of the superfluid neutrons and superconductive protons on the  neutron star core (upper limit of $T$).

It is obvious from Fig.~10 that the majority of the stars lie
outside  the instability windows  predicted by the present
model. There are four  exceptions,  that is the {\it 4U 1608-52},
the {\it SAX J1750.8-2900}, the {\it 4U-1636-536} and the {\it MXB 1658-298}. Obviously, the above stars lie inside the instability window for the case of neutron star mass  $M=1.4 M_{\odot}$ and $M=1.8
M_{\odot}$. The present results are comparable but not similar with those
in Ref.~\cite{Wen-012} and  in contradiction with those presented
in Refs.~\cite{Vidana-012} and ~\cite{Haskell-012}.  More precisely, in Ref.~\cite{Wen-012} the authors employed a simple phenomenological model of a neutron star with a perfectly rigid crust, as mentioned in the Introduction and  they concluded that, at least,  for the case of low neutron star mass $M=1.4 M_{\odot}$, a softer EOS increases  the lower frequency bound of the instability window and therefore the EOSs characterized by $L \leq 65$ MeV are more consistent with the observation of various LMXBs  frequencies.  The model employed  in the present work predicts an even larger instability region, compared to Ref.~\cite{Wen-012} both for low and high values of neutron star mass. This is the reason that even for the case  $M=1.4 M_{\odot}$ four neutron stars lie inside the instability window in contrast  with the finding of Ref.~\cite{Wen-012}. In any case, the main conclusion of the two models is summarized as follows: the stiffness of the equation of state has a strong effect on the width of the instability window and this effect is more pronounced for high values of  the neutron star mass.

On the other hand the author in Ref.~\cite{Vidana-012}, using various microscopic and phenomenological approaches,   found that the r-mode instability region is smaller for those models which give larger values of $L$. The explanation is related with the $L$ dependence of the bulk and the shear viscosities. In particular, the author found that both bulk and shear viscosities increase with $L$ and consequently, damping of the r-mode is more efficient for models with larger $L$. In the present work,  I consider that, in contradiction with the study of Ref.~\cite{Vidana-012},  the damping mechanism is due only  to viscous dissipation at the boundary layer of perfectly rigid crust and fluid core (the study in Ref.~\cite{Wen-012}, is based in  the same consideration). The relative viscosity (given by Eqs.~(\ref{eta-ee-1}) and (\ref{eta-nn-1})) is fixed to the crust-core interface region and is  proportional to the transition density. Thus, it is obvious  by using relation (\ref{nt-L-1}) as well, that the viscosity is a decreasing function of $L$.  The $L$ dependence of the viscosity is well reflected on the values of the fiducial time scales $\tilde{\tau}_{ee}$ and $\tilde{\tau}_{nn}$ (see  Eq.(\ref{tv-1}) and consequently on the values of the critical angular momentum and frequency (according to Eq.(\ref{Omega-c-1})) as exhibited  in Fig.~10.

Moreover, in   \cite{Haskell-012}
the authors by employing the "minimal model", found that a
significant number of systems is well inside the instability
window. A possible explanation of the mentioned contradiction is
the viscous dissipation at the boundary layer between crust and
core, which   is taken into account explicitly in the present work.
Actually, when the authors in Ref.~\cite{Haskell-012} include the
above dissipation mechanism via the "slip" parameter $S$ \cite{Glampedakis-06}
(related with the rigidity of the crust) the majority of the stars {\it
shifted} to the stability area in accordance with the observations,
since the majority of the LMXBs should be out of the instability
window (see also Refs.~\cite{Levin-99,Bondarescu-07}).

I extend also my study on the effect of the isoscalar part of the EOS
to the critical rotational frequency $\nu_c$. The incompressibility $K$, which is one of the main  quantities related directly with the isoscalar
behavior of the EOS, is defined as
\begin{equation}
K=\left. 9n^2\frac{d^2(E/A)}{d^2n}\right|_{n=n_0} \ ({\rm MeV}), \label{inco-1}
\end{equation}
where $E/A$ is the energy per particle of symmetric nuclear matter and $n_0$ the saturation density.
The parametrization of $E/A$ is given in Table~2 of Ref.~\cite{Prakash-97}.
In particular, I kept fixed the nuclear symmetry dependence, and varied the incompressibility in a large range of $K=180-240$ MeV according to the method presented in Ref.~\cite{Prakash-97}. In this case, the energy per particle of symmetric nuclear matter is given by Eq.~(\ref{e-T0}) and
the symmetry energy  by the simple analytical formula
\begin{equation}
E_{sym}(u)=13u^{2/3}+17u \ ({\rm MeV}).
\label{Esym-fu}
\end{equation}
The value of the slope parameter $L$, corresponding to the above ansatz of the symmetry energy,  is $L=77$ MeV.
The results are
presented in Fig.~11. A stiffer   equation of state (higher values
of $K$) leads to lower values of $\nu_c$. The critical frequency
decreases  by around $7\%$ for $K=120$ MeV to $K=240$ MeV. Obviously,
the effect of $K$ on $\nu_c$
is moderate, compared to the effect of the slope parameter $L$, but
not negligible. However, since the most experimental values of $K$
are well constrained around the value $240$ MeV, the uncertainty
related with the isoscalar effects on the equation of state and
consequently on the critical frequency is limited.

The equation of state affects not only the conditions for the
r-mode instability but also the angular velocity evolution of a
neutron star according to Eq.~(\ref{dOmegadt-1}). Actually, the
quantities $Q$ and $\tau_{GR}$ depend mainly on the density
distribution  as well as on the bulk neutron star  properties as the mass and
radius. I examine the approximated case of the frequency evolution considering thermal stability. More precisely I consider a neutron star with mass $M=1.4 M_{\odot}$, temperature $T=8\cdot 10^8$ K, initial frequency $\nu_{in}=700$ Hz and    r-mode amplitude
$\alpha=2\cdot 10^{-7}$. The results are presented in
Fig.~12. It is worth  pointing out that the value of $Q$
moderately depends on the use of a specific  equation of state and varies in
the interval  $Q=0.0945-0.0977$. The equation of state affects
mainly the time scale $\tau_{GR}$. In  Fig.~12(a) is displayed  the
time  evolution of the frequency of a neutron star for the five
EOSs. The EOS effects
are more pronounced in the case of the rate of the frequency
$d\nu/dt$ as indicated in Fig.~12(b). A stiffer EOS leads to
a higher value of the spindown rate. Only when the frequency
approachec the critical value, the spindown rate appears to be model
independent.

In Fig.~12(c) I plot the dependence of the spindown rate $d\nu/dt$ on the frequency. For the same values of the frequency a higher value
of $L$ leads to a larger value of the spindown rate (the neutron
star approaches its critical frequency faster). It is
worthwhile to notice that the results presented in Fig.~12(c) are
sensitive both to the core temperature and mainly to the values of
the r-mode amplitude $\alpha$.   The most interesting feature of
this plot is related with the range covered by the implied
equation of states. Thus,  in the same figure I include the observed
spin-down rate for three cases (IGR J00291+5934, XTE J1751-305 and
SAX J1808-3658) in
comparison with the theoretical predictions~\cite{Mahmoodifar-013,Alford-013,Manchester-05,Paturno-010,Paturno-012a,Paturno-012b}.  I estimate that it
may be possible, by a suitable treatment of  observations  and
related theoretical predictions of  the spin frequency and
spindown rate of known neutron star,  to impose  additional
constraints on the nuclear equation of state. However, it is
necessary to proceed with a  more detailed calculation concerning
the r-mode instability and evolution. In order to establish the above statement is necessary to use more elaborate
equations of state (with additional degrees of freedom), to
consider additional  dissipation mechanisms  and treat
more carefully  the thermal evolution.

It should be noted that in the present analysis additional degrees of
freedom, like quark and hyperon matter are not considered for the
construction of the equation of state. It is known that, in
most cases, the presence of quark and hyperon matter affect
appreciably the equation of state by softening the density
dependence mainly for high densities. This means that the bulk
properties of neutron stars will also be affected and consequently
the time scales of the r-mode will be affected too. In addition, the
presence of quark and hyperons infuence the dissipation mechanisms
since one has to take into account the shear and also the bulk
viscosities due to the presence of this kind of matter. Actually,
there are several recent studies in this
direction~\cite{Alford-2012,Haskell-012,Andersson-010,Haskell-07,Alford-2012b,Linddblom-02,Rupak-013,Chatterjee07}.
In any case, when more degrees of freedom are taken into account,
the analysis becomes more complete and consequently  more reliable.

In view of the above discussion, I
consider that another issue worth examining is the connection
between the observed neutron stars in low-mass X-ray binaries and
the nuclear physics input via the equation of state~\cite{Read-09-a,Read-09-b,Rezzolla-014}. It  would be very interest, if it is possible to constrain the nuclear
physics input (for example the slope parameter $L$) employing the
related observation data. In general, this is a very complex problem,
since the nuclear equation of state affects in different ways the
r-mode instability and consequently additional
work is needed as well as one has to illustrate further this point. However
additional theoretical and observational work  must be dedicated
before  being  able to impose strong constraints on the
implemented  EOSs.

Finally, it is well known that for high temperature $T > 10^{10}$
K, which characterize mainly a newborn neutron star, the bulk
viscosity is the dominant dissipation mechanism. However such a
kind of dissipation is not considered in the present work.
Actually, at high temperature, in a self-consistent treatment one
has to consider also temperature effects  on the nuclear equation
of state. It is known \cite{Xu-2010} that temperature influences
not only the density dependence of the equation of state but in
addition both the transition density and the proton fraction.
Actually the present MDIM model can be extended to include also
the temperature effect on equation of state
\cite{Moustakidis-07-1,Moustakidis-07-2,Moustakidis-08,Moustakidis-09-1,Moustakidis-09-2}.
Work along these lines is in progress.

%%%%%%%%%%%%%%%%%%%%%%%%%
\section{Summary}
%%%%%%%%%%%%%%%%%%%%%%%%%
In the present work I consider the effect, on r-mode
instability, due to the presence of a solid crust in a neutron
star. By employing a phenomenological nuclear model I calculated
the equation of state of $\beta$-stable matter which characterizes
the neutron star core and is used for the location of the
transition density at the inner edge between the liquid core and
the solid crust. The stiffness of the equation of state
parameterized via the slope parameter $L$, was varied on the
interval $72.5\ {\rm MeV} \leq L \leq 110$ MeV. The gravitational
and the viscous time scales depend directly on the parameter $L$
as well as indirectly on the transition density $n_t$. As a
consequence the critical angular velocity as well as the critical
temperature depend on the equation of state. I found also  that
the instability window drops by $\simeq 20-40 $ Hz when the mass
of a neutron star is raised from $M=1.4 M_{\odot}$ to $M=1.8
M_{\odot}$ and also that the use of a stiffer equation of state
increases the instability window. I
compared the r-mode instability window for the five selected
equation of states with those of the observed neutron stars in
LMXBs and MSRPs for $M=1.4 M_{\odot}$ and $M=1.8 M_{\odot}$. I
found that the majority of the stars lie outside of the
instability windows.  Finally, I estimated the time
evolution of the spin frequency and spin-down rate for the
selected EOS's for a $M=1.4 M_{\odot}$ neutron star mass in
comparison with three observed cases. I conclude  that it may be
possible to impose additional constraints on the nuclear equation
of state, by a suitable combination of  observations  and relative
theoretical predictions.

%%%%%%%%%%%%%%%%%%%%%%%%%%%%%
\section*{Acknowledgments}
%%%%%%%%%%%%%%%%%%%%%%%%%%%%%%
This work was supported by the German Science Council (DFG) via
SFB/TR7 and by the Aristotle University of Thessaloniki Research Committee under Contract No. 89286. The author would like to thank the Theoretical
Astrophysics Department of the University of Tuebingen, where part
of this work was performed  and Professor K. Kokkotas  for his
useful comments on the preparation of the manuscript. The author
thanks Dr. C.P. Panos for his remarks on the present paper and Dr. A. Mytidis for useful discussions.

%%%%%%%%%%%%%%%%%%%%%%%%%%%%%%%%%%%%%%%%%%%%%%%%%%%%%%%%%%%%%%%%%%%%%%
%FIGURE-1
\begin{figure}
\centering
\includegraphics[height=7.1cm,width=8.1cm]{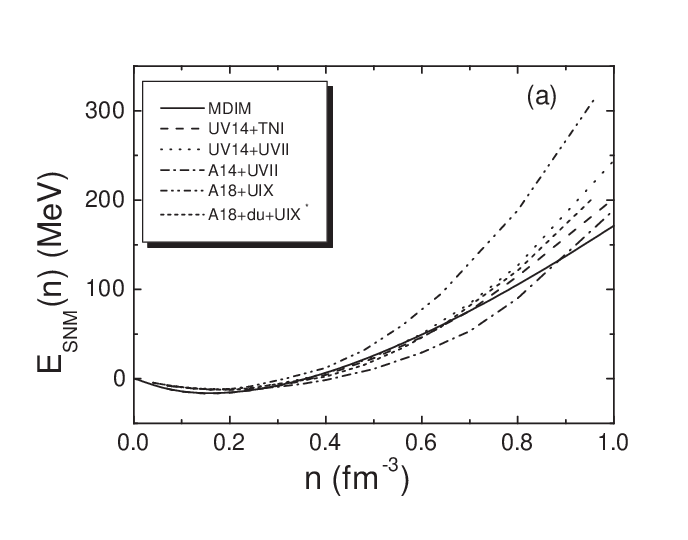}\
\includegraphics[height=7.1cm,width=8.1cm]{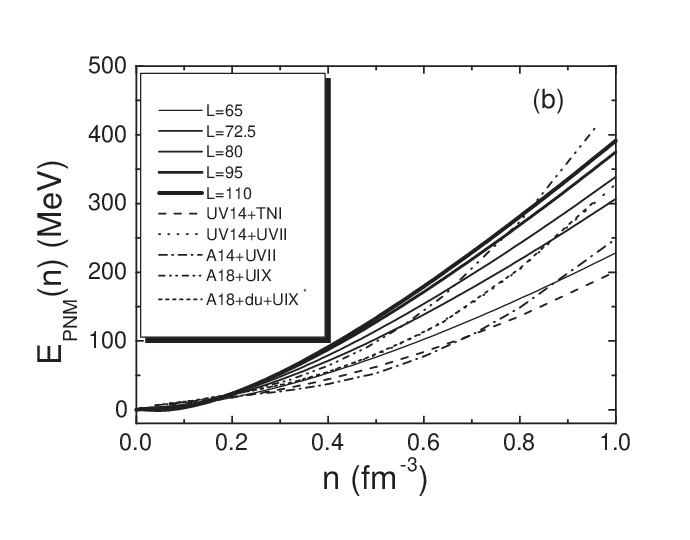}\
\caption{ The energy per baryon of symmetric nuclear matter (a)
and pure neutron matter (b), as a function of the baryon density
$n$, of the present model (MDIM) in comparison with those
originating from realistic calculations. More details for the
models UV14+TNI, UV14+UVII and  AV14+UVII in
Ref.~\cite{Wiringa-88-b} and for the models A18+UIX and
A18+du+UIX$^*$ in Ref.~\cite{Pandharipande-98}. } \label{Esnm-pnm}
\end{figure}
%%%%%%%%%%%%%%%%%%%%%%%%%%%%%%%%%%%%%%%%%%%%%%%%%%%%%%%%%%%%%%%%%%%%%%
%%%%%%%%%%%%%%%%%%%%%%%%%%%%%%%%%%%%%%%%%%%%%%%%%%%%%%%%%%%%%%%%%%%%%%
%%%%%%%%%%%%%%%%%%%%%%%%%%%%%%%%%%%%%%%%%%%%%%%%%%%%%%%%%%%%%%%%%%%%%%

%FIGURE-2
\begin{figure}
\centering
\includegraphics[height=7.1cm,width=8.1cm]{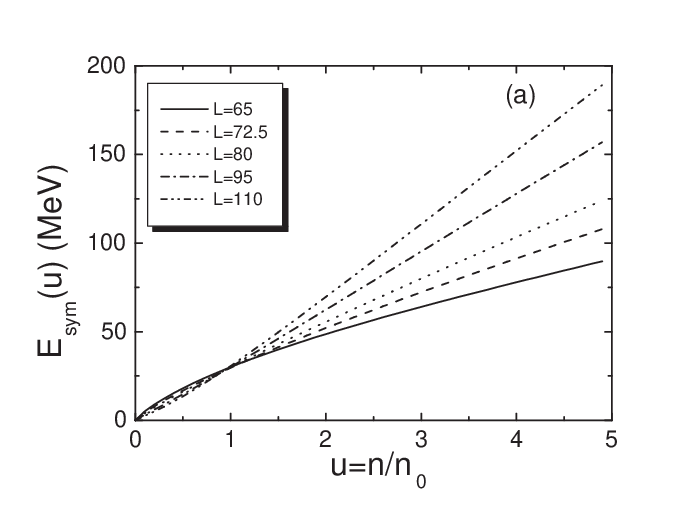}\
\includegraphics[height=7.1cm,width=8.1cm]{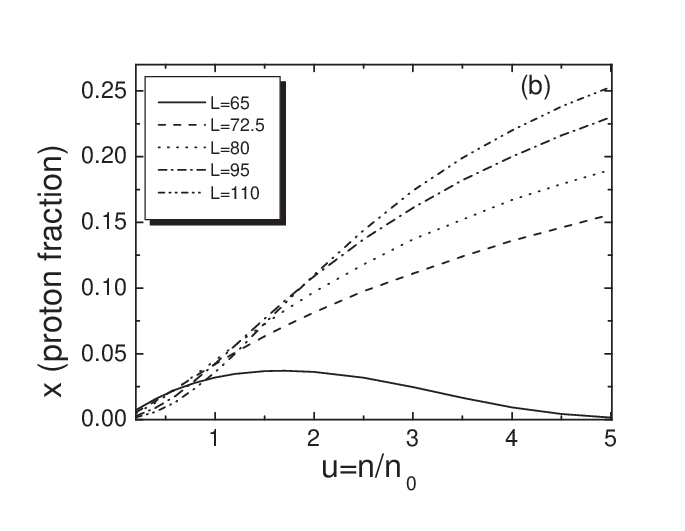}\
\includegraphics[height=7.1cm,width=8.1cm]{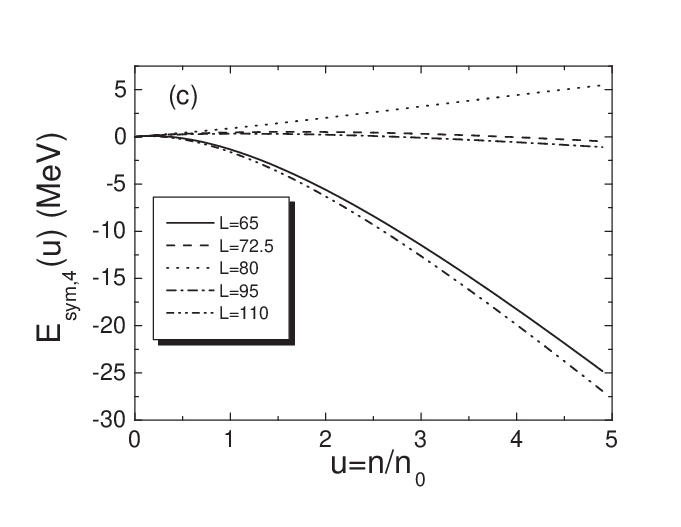}\
\caption{ The nuclear symmetry energy (a) the proton fraction
(b) and the fourth order  term $E_{sym,4}(u)$ of  the symmetry energy (c) as a function of the ratio $u=n/n_0$ for various values of the
slope parameter $L$. } \label{}
\end{figure}
%%%%%%%%%%%%%%%%%%%%%%%%%%%%%%%%%%%%%%%%%%%%%%%%%%%%%%%%%%%%%%%%%%%%%%
%%%%%%%%%%%%%%%%%%%%%%%%%%%%%%%%%%%%%%%%%%%%%%%%%%%%%%%%%%%%%%%%%%%%%%

%%%%%%%%%%%%%%%%%%%%%%%%%%%%%%%%%%%%%%%%%%%%%%%%%%%%%%%%%%%%%%%%%%%%%%

%FIGURE-3
\begin{figure}
\centering
\includegraphics[height=7.1cm,width=8.1cm]{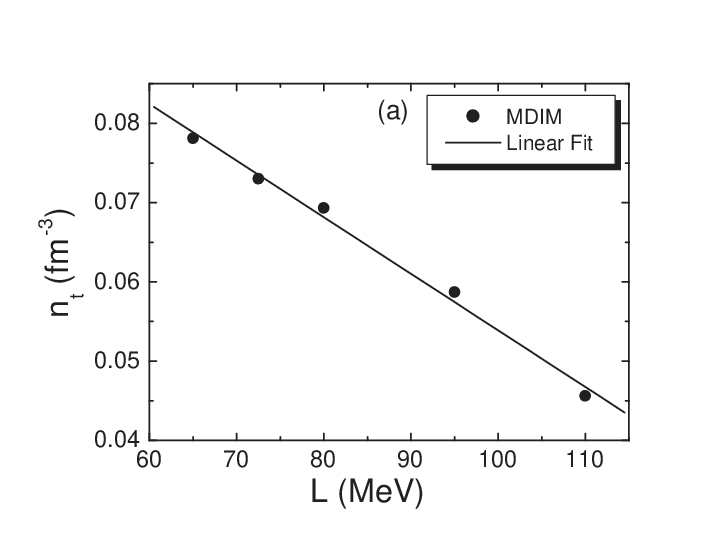}\
\includegraphics[height=7.1cm,width=8.1cm]{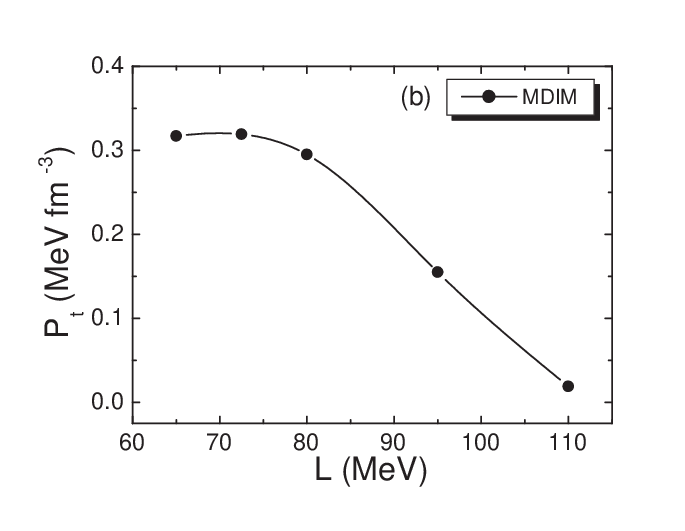}\
\caption{The transition baryon density $n_t$ (a) and  the
transition pressure $P_t$ (b) as a function of the slope parameter
$L$. For more details about the linear fit see text. } \label{}
\end{figure}
%%%%%%%%%%%%%%%%%%%%%%%%%%%%%%%%%%%%%%%%%%%%%%%%%%%%%%%%%%%%%%%%%%%%%%
%%FIGURE-4
\begin{figure}
\centering
\includegraphics[height=7.1cm,width=8.1cm]{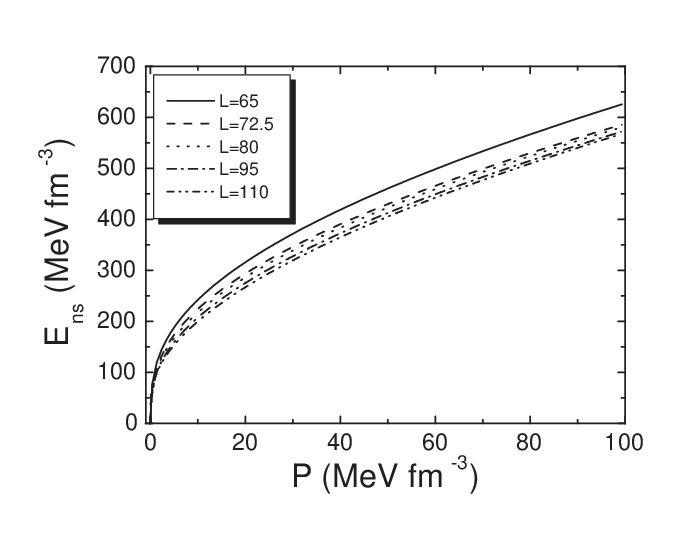}\
\caption{The energy density $E_{ns}$, taking into account both the fluid core and the solid crust matter, of neutron star matter as a function of the pressure $P$ for various values of the
slope parameter $L$. The data used for densities lower than the transition density  $n_t$ have been taken from \cite{Feynman-49} and \cite{Baym-71} (for more details see text).}  \label{}
\end{figure}
%%%%%%%%%%%%%%%%%%%%%%%%%%%%%%%%%%%%%%%%%%%%%%%%%%%%%%%%%%%%%%%%%%%%%%

%%%%%%%%%%%%%%%%%%%%%%%%%%%%%%%%%%%%%%%%%%%%%%%%%%%%%%%%%%%%%%%%%%%%%%
%FIGURE-5
\begin{figure}
\centering
\includegraphics[height=9.1cm,width=8.1cm]{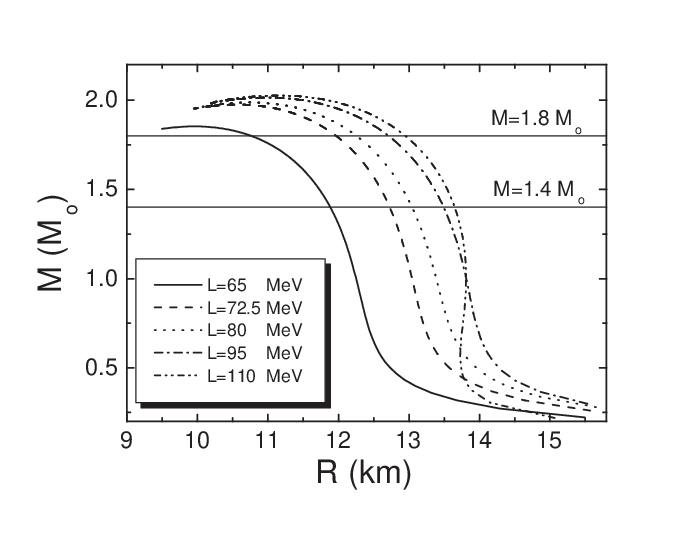}\
\caption{The mass-radius relations for the selected EOSs }
\label{}
\end{figure}
%%%%%%%%%%%%%%%%%%%%%%%%%%%%%%%%%%%%%%%%%%%%%%%%%%%%%%%%%%%%%%%%%%%%%%

%%%%%%%%%%%%%%%%%%%%%%%%%%%%%%%%%%%%%%%%%%%%%%%%%%%%%%%%%%%%%%%%%%%%%%
%FIGURE-6
\begin{figure}
\centering
\includegraphics[height=8.1cm,width=8.1cm]{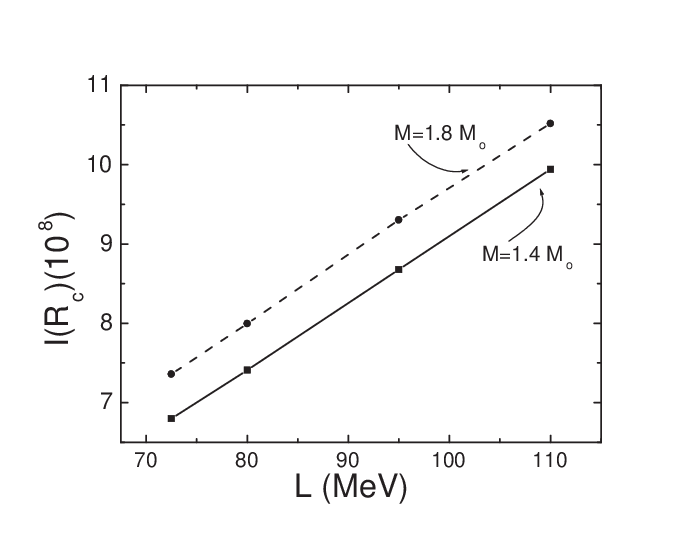}\
\caption{The values of the fundamental integral $\displaystyle
I(R_c)=\int_{0}^{R_c} \epsilon (r) r^6 dr$ as a function of the
slope parameter $L$ for the interval $72.5 \ {\rm MeV} \leq L \leq
110 \ {\rm MeV}$. For more details about the linear fit see text.}
\label{fig-L-Int}
\end{figure}
%%%%%%%%%%%%%%%%%%%%%%%%%%%%%%%%%%%%%%%%%%%%%%%%%%%%%%%%%%%%%%%%%%%%%%
%%%%%%%%%%%%%%%%%%%%%%%%%%%%%%%%%%%%%%%%%%%%%%%%%%%%%%%%%%%%%%%%%%%%%%
%FIGURE-7
\begin{figure}
\vspace{3cm} \centering
\includegraphics[height=8.1cm,width=8.1cm]{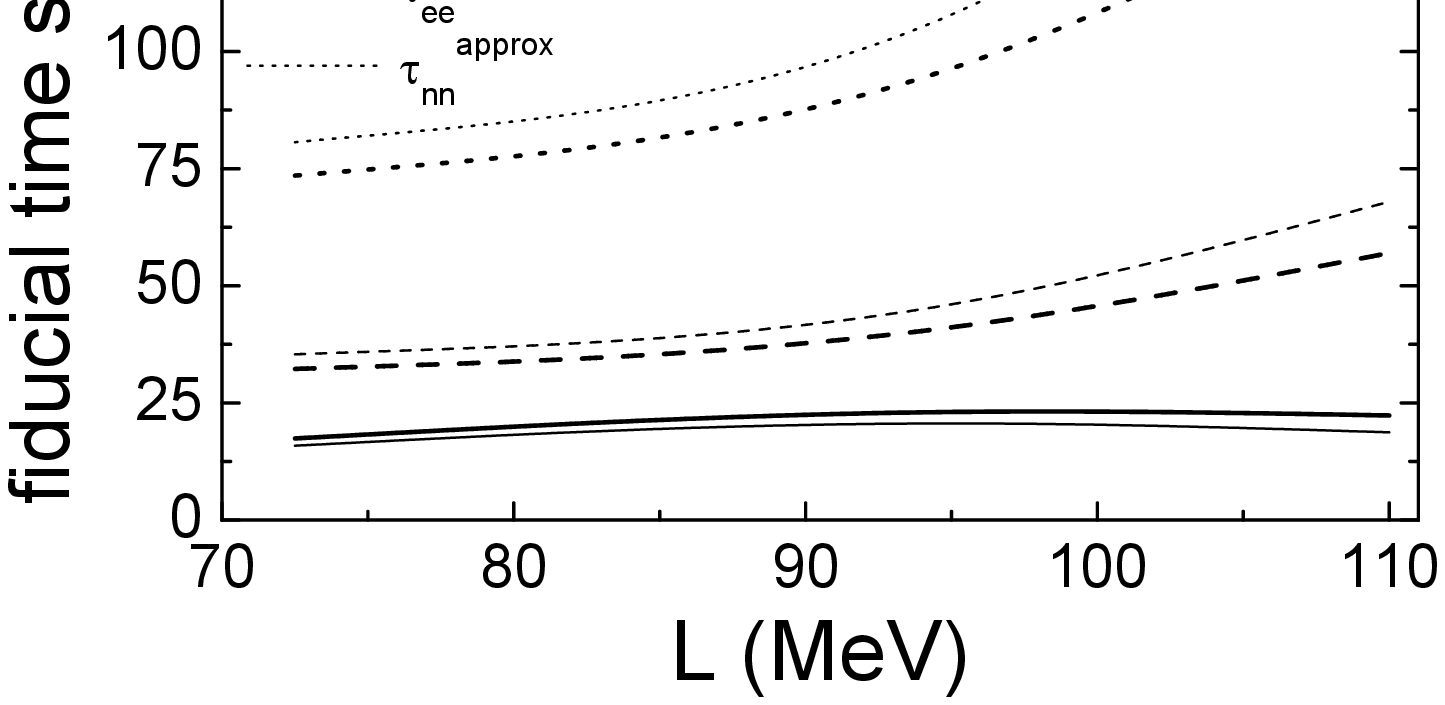}\
\includegraphics[height=8.1cm,width=8.1cm]{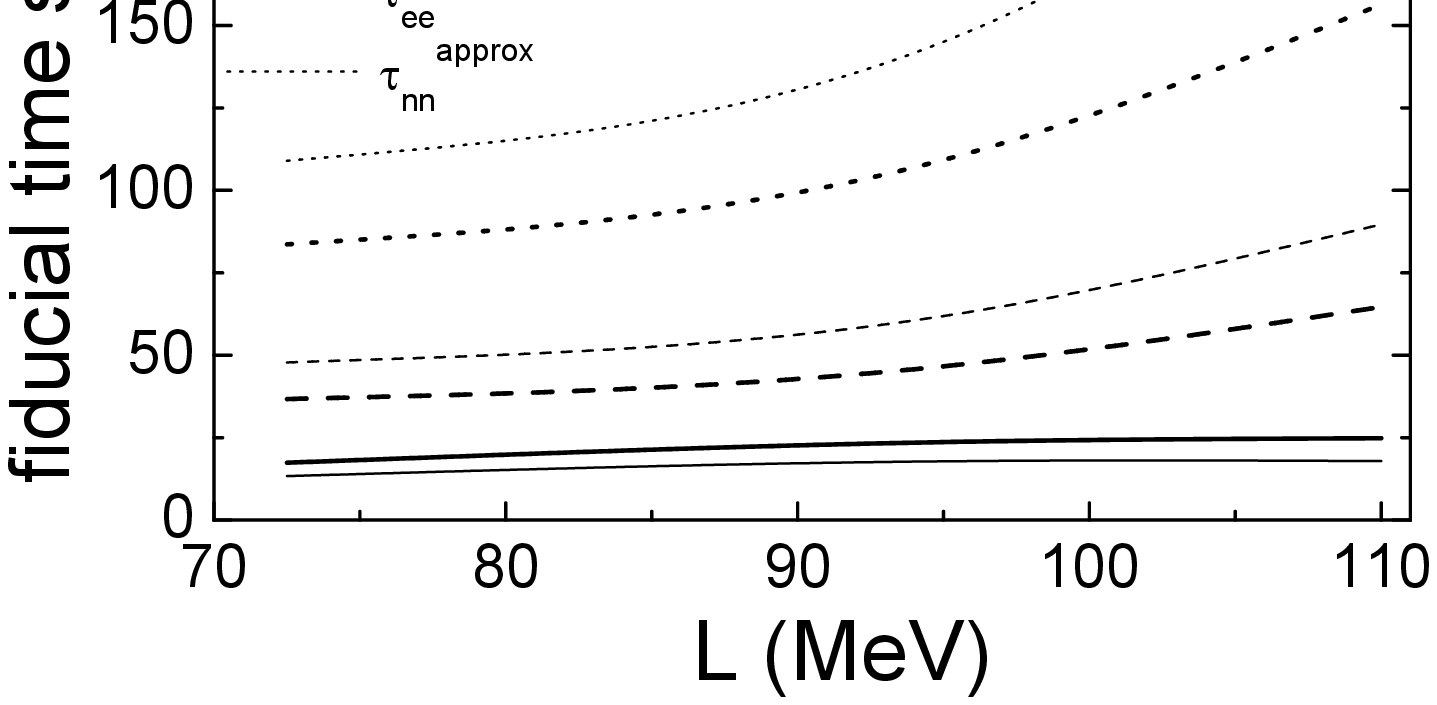}\
\vspace{-3.5cm} \caption{The fiducial time scales
$\tilde{\tau}_{GR}$, $\tilde{\tau}_{ee}$, $\tilde{\tau}_{nn}$ as
well as $\tilde{\tau}_{GR}^{approx}$,
$\tilde{\tau}_{ee}^{approx}$, $\tilde{\tau}_{nn}^{approx}$ as a
function of the slope parameter $L$ for a neutron star with mass
$M=1.4 M_{\odot}$ (a) and $M=1.8 M_{\odot}$ (b). In case  $M=1.4
M_{\odot}$ the gravitational times scales are multiplied by a
factor 5 and in case $M=1.8 M_{\odot}$ are multiplied by a factor
20.} \label{tgr-nn-ee-L}
\end{figure}
%%%%%%%%%%%%%%%%%%%%%%%%%%%%%%%%%%%%%%%%%%%%%%%%%%%%%%%%%%%%%%%%%%%%%%
%%%%%%%%%%%%%%%%%%%%%%%%%%%%%%%%%%%%%%%%%%%%%%%%%%%%%%%%%%%
%%%%%%%%%%%%%%%%%%%%%%%%%%%%%%%%%%%%%%%%%%%%%%%%%%%%%%%%%%%%
%%%%%%%%%%%%%%%%%%%%%%%%%%%%%%%%%%%%%%%%%%%%%%%%%%%%%%%%%%%%%%%%%%%%%%

%FIGURE-8
\begin{figure}
\vspace{3cm} \centering
\includegraphics[height=8.1cm,width=8.5cm]{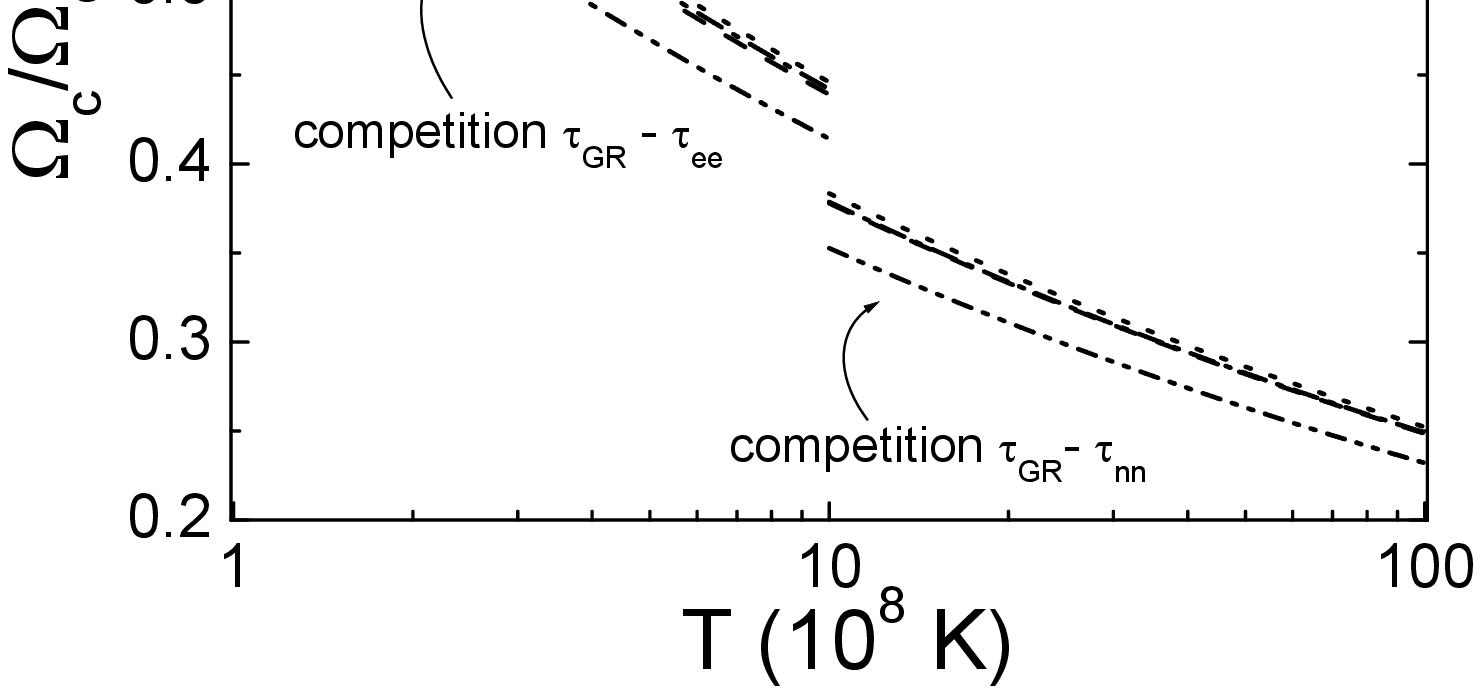}\
\includegraphics[height=8.1cm,width=8.5cm]{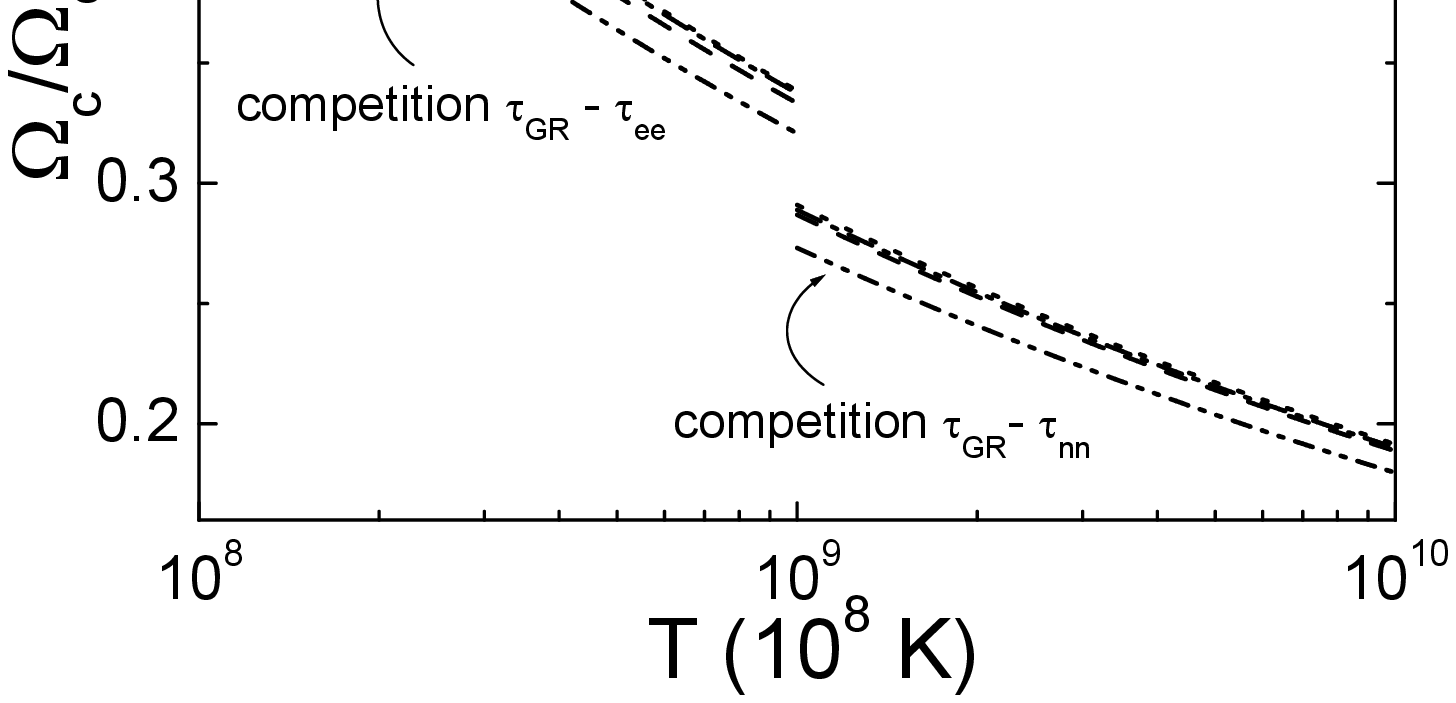}\
\vspace{-3.5cm} \caption{Temperature dependence of the critical
angular velocity ratio $\Omega_c/\Omega_0$ for a neutron star with
mass $M=1.4 M_{\odot}$ (a) and $M=1.8 M_{\odot}$ (b) constructed
for the selected EOSs. } \label{}
\end{figure}
%%%%%%%%%%%%%%%%%%%%%%%%%%%%%%%%%%%%%%%%%%%%%%%%%%%%%%%%%%%%%%%%%%%%%%
%%%%%%%%%%%%%%%%%%%%%%%%%%%%%%%%%%%%%%%%%%%%%%%%%%%%%%%%%%%%%%%%%%%%%%
%%%%%%%%%%%%%%%%%%%%%%%%%%%%%%%%%%%%%%%%%%%%%%%%%%%%%%%%%%%%%%%%%%%%%%
%%%%%%%%%%%%%%%%%%%%%%%%%%%%%%%%%%%%%%%%%%%%%%%%%%%%%%%%%%%%%%%%%%%%%%

%FIGURE-9
\begin{figure}
\centering
\includegraphics[height=8.1cm,width=8.5cm]{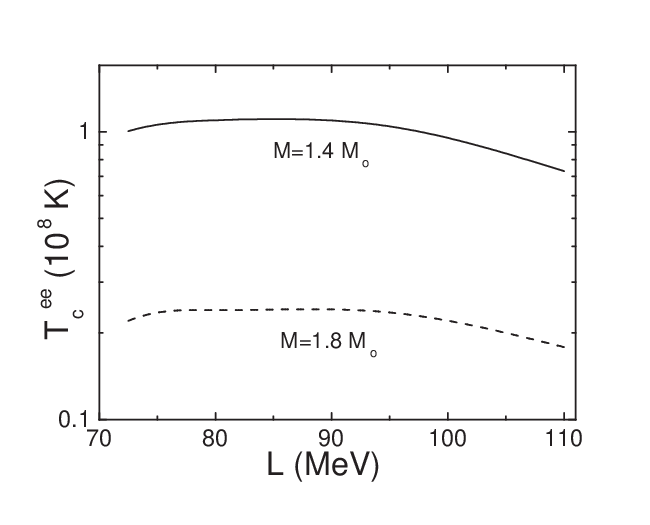}\
\caption{The critical temperature as a function of  the slope
parameter $L$   for a
neutron star with mass $M=1.4 M_{\odot}$ and $M=1.8 M_{\odot}$.
} \label{}
\end{figure}
%%%%%%%%%%%%%%%%%%%%%%%%%%%%%%%%%%%%%%%%%%%%%%%%%%%%%%%%%%%%%%%%%%%%%%

%%%%%%%%%%%%%%%%%%%%%%%%%%%%%%%%%%%%%%%%%%%%%%%%%%%%%%%%%%%%%%%%%%%%%%
%FIGURE-10
\begin{figure}
\vspace{3cm} \centering
\includegraphics[height=8.1cm,width=8.6cm]{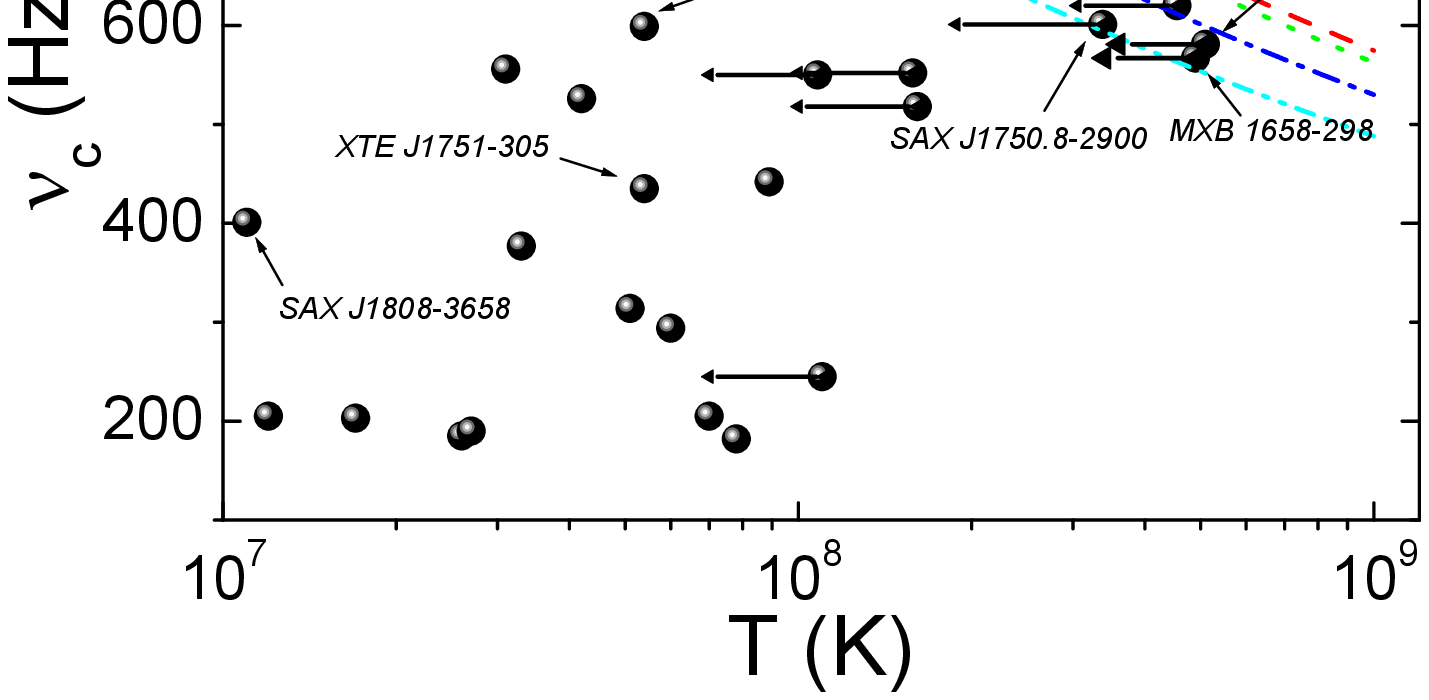}\
\includegraphics[height=8.1cm,width=8.6cm]{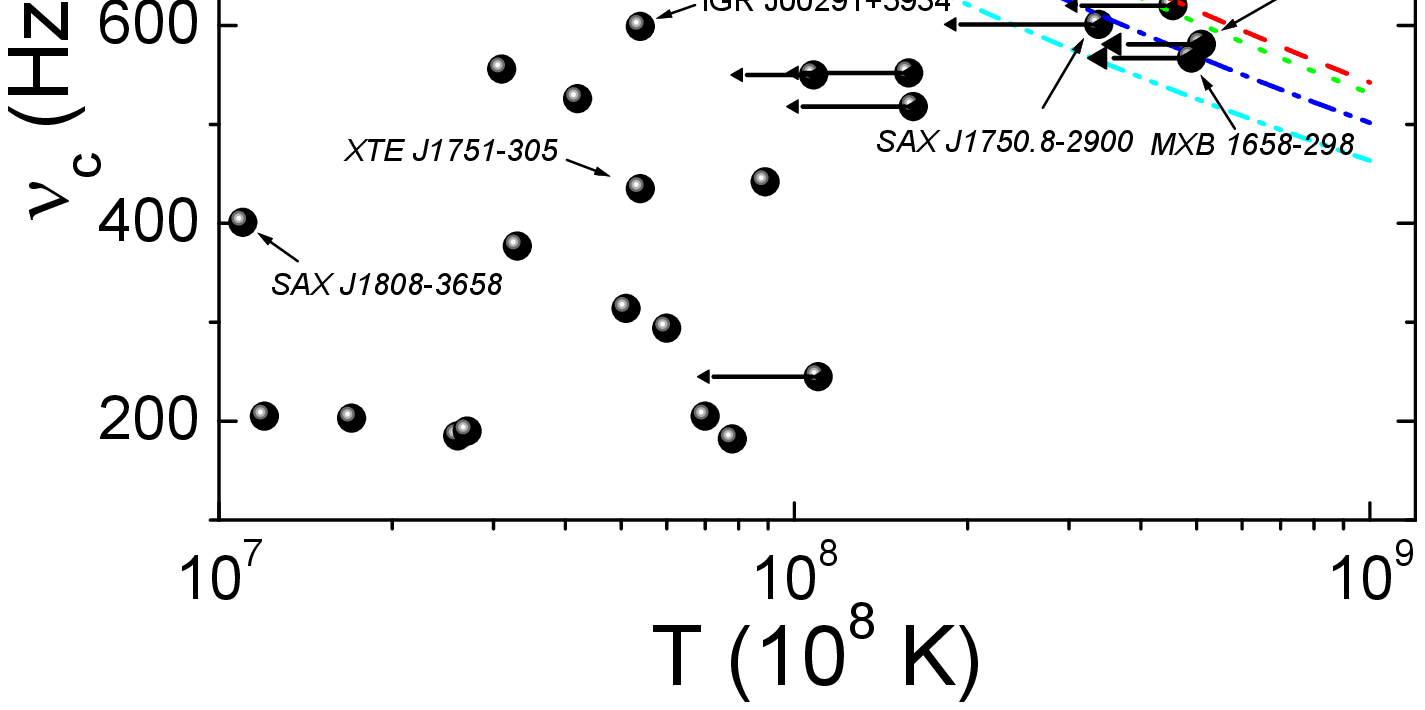}\
\vspace{-3.5cm} \caption{(Color online) The critical frequency  temperature
dependence for  a neutron star with mass $M=1.4 M_{\odot}$ (a) and
$M=1.8 M_{\odot}$ (b) constructed for the selected EOSs. The
observed cases of LMXBs and MSRPs from Haskell et al. \cite{Haskell-012},  are also included for a
comparison. The horizontal vectors extending leftward exhibit the uncertainties of the core temperature.  The cases {\it IGR J00291+5934}, {\it XTE J1751-305} and {\it SAX J1808-3658} with well known observation  spindown rate, are also indicated (see also Fig.12(c)).}  \label{frequanecy-T-M18}
\end{figure}
%%%%%%%%%%%%%%%%%%%%%%%%%%%%%%%%%%%%%%%%%%%%%%%%%%%%%%%%%%%%%%%%%%%%%%

%%%%%%%%%%%%%%%%%%%%%%%%%%%%%%%%%%%%%%%%%%%%%%%%%%%%%%%%%%%%%%%%%%%%%
%FIGURE-11
\begin{figure}
\vspace{3cm} \centering
\includegraphics[height=8.1cm,width=8.1cm]{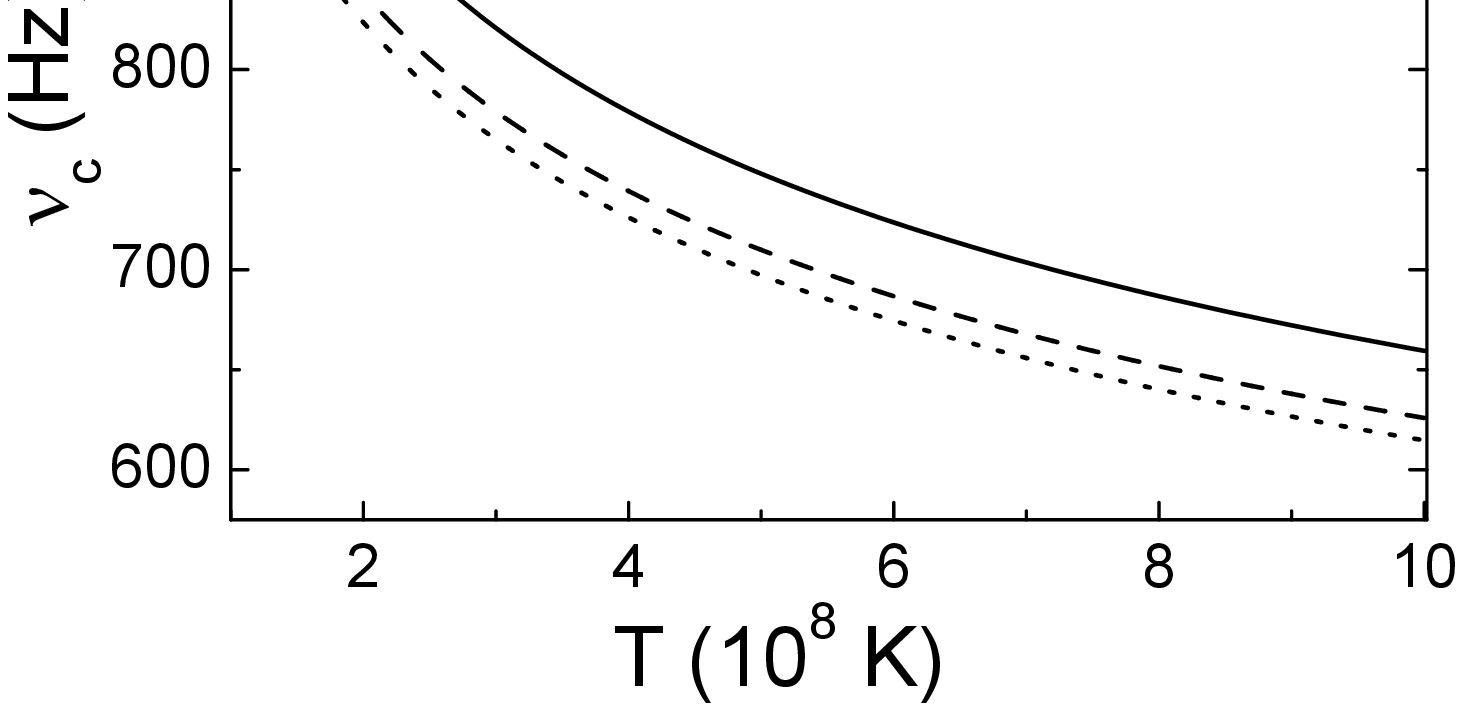}\
\vspace{-3.5cm} \caption{The critical frequency  temperature
dependence for  a neutron star with mass $M=1.4 M_{\odot}$
 for a fixed $E_{sym}(u)$, given by Eq.~(\ref{Esym-fu}) and for three values of the incompressibility $K$.  }  \label{frequanecy-T-M18}
\end{figure}
%%%%%%%%%%%%%%%%%%%%%%%%%%%%%%%%%%%%%%%%%%%%%%%%%%%%%%%%%%%%%%%%%%%%%%

%%%%%%%%%%%%%%%%%%%%%%%%%%%%%%%%%%%%%%%%%%%%%%%%%%%%%%%%%%%%%%%%%%%%%
%FIGURE-12
\begin{figure}
\vspace{3cm} \centering
\includegraphics[height=8.1cm,width=8.1cm]{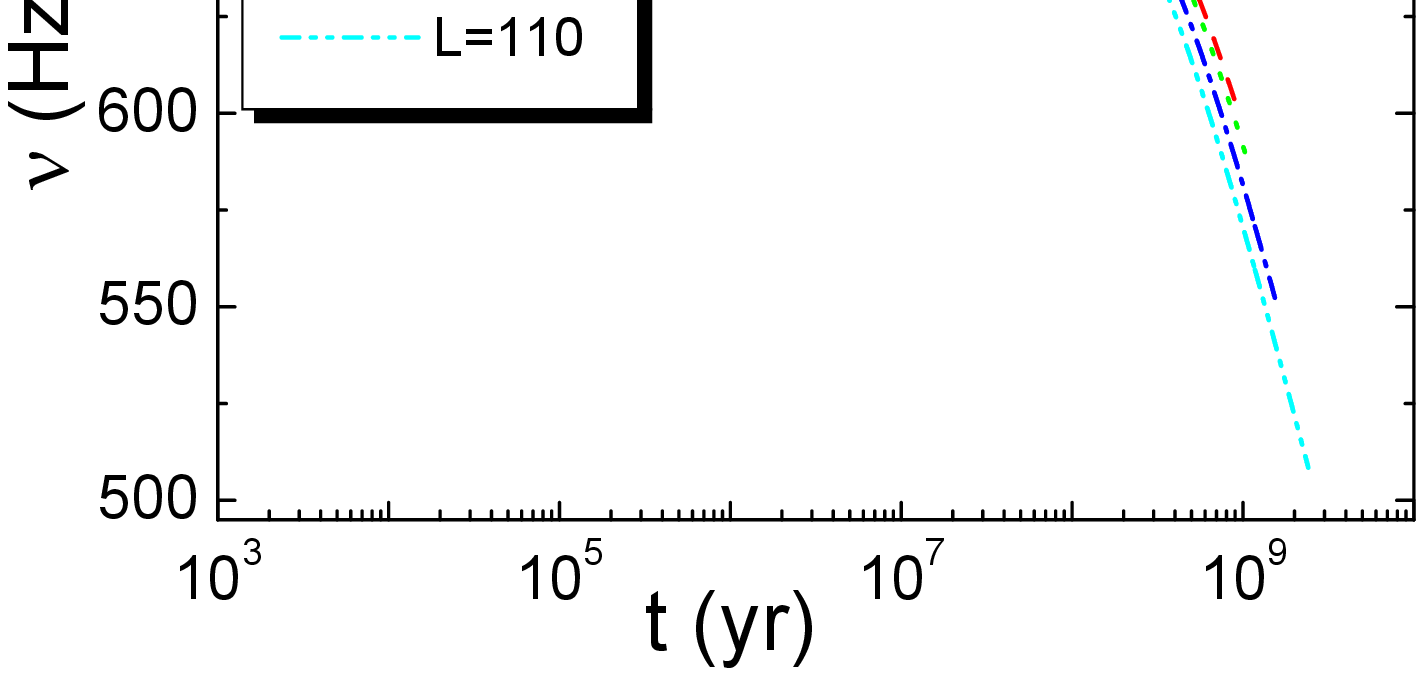}\
\includegraphics[height=8.1cm,width=8.1cm]{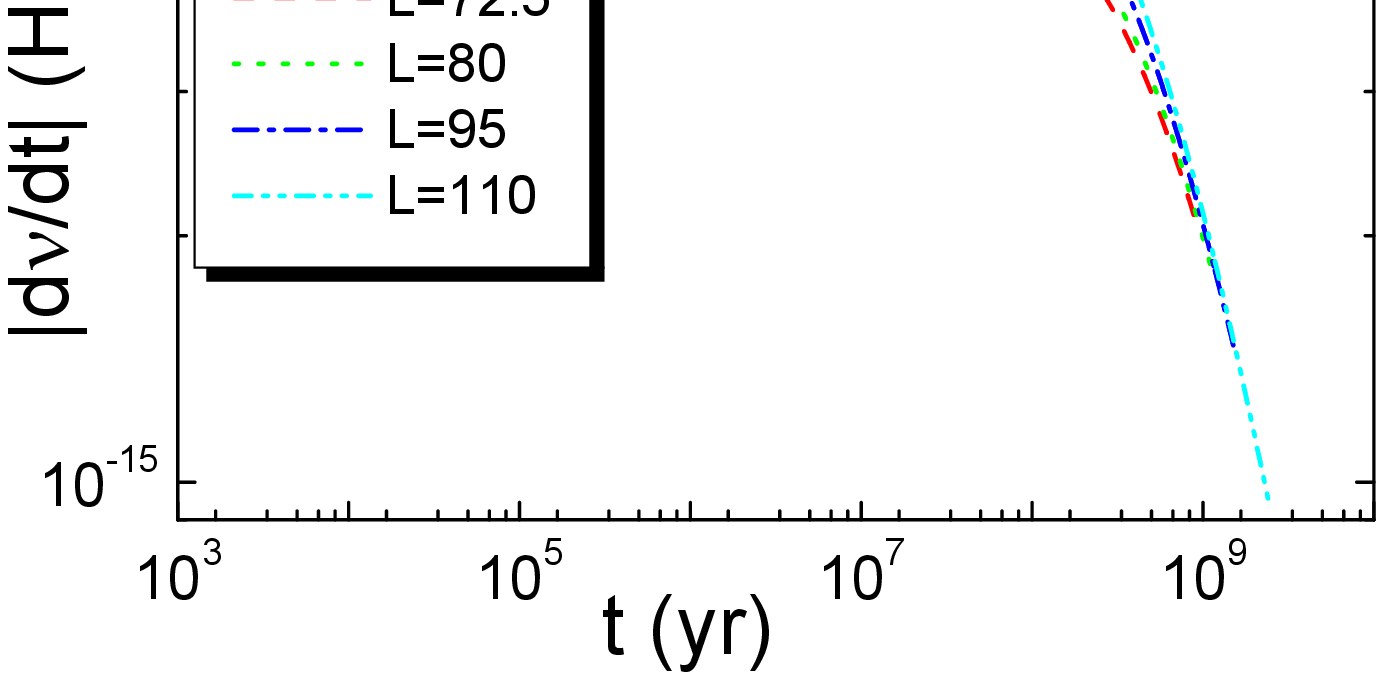}\
\includegraphics[height=8.1cm,width=8.1cm]{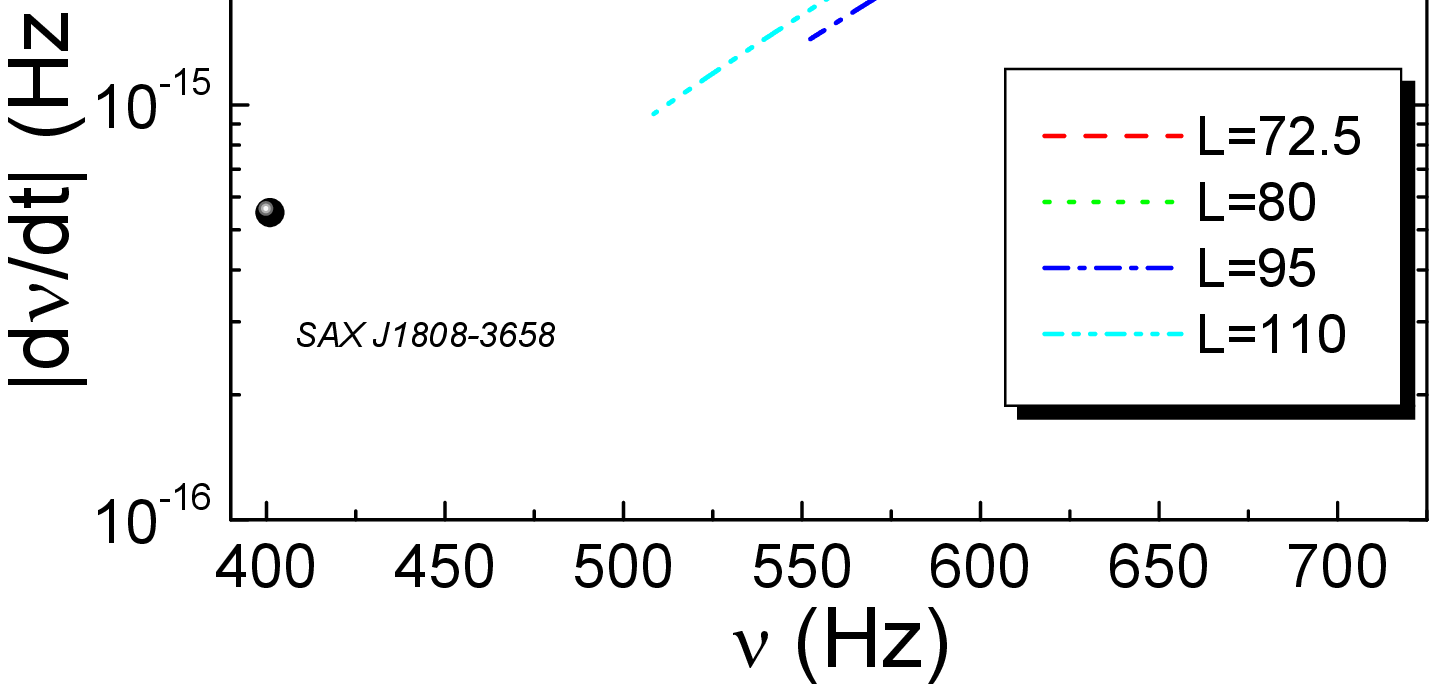}\
\vspace{-3.5cm} \caption{(Color online)  (a) The time evolution of the spin frequency, (b)  the spin-down rate evolution, for  a neutron star with mass $M=1.4 M_{\odot}$, for the selected EOSs and  (c) the spin-down rate versus the spin frequency, for the selected EOSs compared to three observed pulsars data \cite{Mahmoodifar-013,Manchester-05}.    }  \label{frequanecy-T-M18}
\end{figure}
%%%%%%%%%%%%%%%%%%%%%%%%%%%%%%%%%%%%%%%%%%%%%%%%%%%%%%%%%%%%%%%%%%%%%%

%\newpage

%%%%%%%%%%%%%%%%%%%%%%%%%%%%%%%%%%%%%%%%%%%%%%%%%%%%%%%%%%%%%%%%%%%%%%
 \begin{table}[h]
\begin{center}
\caption{The slope parameter $L$ (in MeV), the transition density
$n_t$ (in fm$^{-3}$), the transition pressure $P_t$ (in MeV
fm$^{-3}$), the total radius $R$ (in km), the core radius $R_c$
(in km), the core mass $M_c$ in $ M_{\odot}$ and the central
pressure $P_c$ (in MeV fm$^{-3}$) correspond to a neutron star
with mass  $M=1.4 M_{\odot}$. }
 \label{t:1}
\vspace{0.5cm}
\begin{tabular}{|c|c|c|c|c|c|c|}
\hline
 $L$     &   $n_t$    & $P_t$    &  $R$        &  $R_c$     &  $M_c$     &  $P_{c}$       \\
\hline
 65      & 0.0781     & 0.317    & 11.885       & 10.956     &   1.383         & 86.4              \\
 \hline
 72.5    &  0.0730    & 0.319    &  12.725     &  11.631    & 1.379           &  63.2             \\
 \hline
 80      & 0.0693     & 0.295    &  13.041      & 11.898      & 1.378          &   57.5           \\
 \hline
 95      & 0.0587     & 0.155    &  13.490      &  12.385    & 1.386         &  49.2           \\
 \hline
 110     & 0.0456     & 0.0188   &  13.646      & 12.805     &  1.397         & 43.5           \\
 \hline
 %\hline
\end{tabular}
\end{center}
\end{table}

%%%%%%%%%%%%%%%%%%%%%%%%%%%%%%%%%%%%%%%%%%%%%%%%%%%%%%%%\newpage
%%%%%
%%%%%%%%%%%%%%%%%%%%%%%%%%%%%%%%%%%%%%%%%%%%%%%%%%%%%%%%%%%%%%%%%%%%%%
 \begin{table}[h]
\begin{center}
\caption{The same as in Table~1 for $M=1.8 M_{\odot}$. }
 \label{t:1}
\vspace{0.5cm}
\begin{tabular}{|c|c|c|c|c|c|c|}
\hline
 $L$     &   $n_t$    & $P_t$    &  $R$        &  $R_c$     &  $M_c$     &  $P_{c}$       \\
\hline
 65      & 0.0781     & 0.317    & 10.757       & 10.287     &   1.793         & 331               \\
 \hline
 72.5    &  0.0730    & 0.319    &  11.965      &  11.328    & 1.788           &  176             \\
 \hline
 80      & 0.0693     & 0.295    &  12.253      & 11.584      & 1.788          &   159           \\
 \hline
 95      & 0.0587     & 0.155    &  12.706      &  12.052    & 1.792         &  133.5           \\
 \hline
 110     & 0.0456     & 0.0188   &  12.951      & 12.444     &  1.798        & 117           \\
 \hline
 %\hline
\end{tabular}
\end{center}
\end{table}

%%%%%%%%%%%%%%%%%%%%%%%%%%%%%%%%%%%%%%%%%%%%%%%%%%%%%%%%\newpage

%%%%%%%%%%%%%%%%%%%%%%%%%%%%%%%%
%%%%%%%%%%%%%%%%%%%%%%%%%%%%%%%%%%%%%%%%%%%%%%%%%%%%%%%%%%%%%%%%%%%%%%
 \begin{table}[h]
\begin{center}
\caption{The fiducial time scales (in sec) for a neutron star with
mass  $M=1.4 M_{\odot}$ }
 \label{t:1}
\vspace{0.5cm}
\begin{tabular}{|c|c|c|c|c|c|c|}
\hline
 $L$     &  $\tilde{\tau}_{GR}$     &  $\tilde{\tau}_{GR}^{approx}$      &    $\tilde{\tau}_{v_{ee}}$   &   $\tilde{\tau}_{v_{ee}}^{approx}$   &   $\tilde{\tau}_{v_{nn}}$    &   $\tilde{\tau}_{v_{nn}}^{approx}$          \\
\hline
  72.5    & -3.484    & -3.176    &  32.254      &  35.380    & 73.526           &  80.656            \\
 \hline
 80      & -3.986     & -3.637    &  33.833      & 37.076      & 77.637          &   85.076          \\
 \hline
 95      & -4.615     & -4.123    &  41.125      &  46.035    & 96.377        &  107.882        \\
 \hline
 110     & -4.468     & -3.748   &  56.977     &  67.936     &  137.840        & 164.353         \\
 \hline
 %\hline
\end{tabular}
\end{center}
\end{table}

%%%%%%%%%%%%%%%%%%%%%%%%%%%%%%%%%%%%%%%%%%%%%%%%%%%%%%%%\newpage

%%%%%%%%%%%%%%%%%%%%%%%%%%%%%%%%%%%%%%%%%%%%%%%%%%%%%%%%%%%%%%%%%%%%%%
 \begin{table}[h]
\begin{center}
\caption{The fiducial time scales (in sec) for a neutron star with
mass  $M=1.8 M_{\odot}$ }
 \label{t:1}
\vspace{0.5cm}
\begin{tabular}{|c|c|c|c|c|c|c|}
\hline
 $L$     &  $\tilde{\tau}_{GR}$     &  $\tilde{\tau}_{GR}^{approx}$      &    $\tilde{\tau}_{v_{ee}}$   &   $\tilde{\tau}_{v_{ee}}^{approx}$   &   $\tilde{\tau}_{v_{nn}}$    &   $\tilde{\tau}_{v_{nn}}^{approx}$          \\
 \hline
 72.5    &  -0.870   &   -0.668  &   36.686     &   47.792   &    83.629        &    108.946         \\
 \hline
 80      &   -0.992   &  -0.760   &  38.415      &  50.145     &   88.150        &   115.068          \\
 \hline
 95      &   -1.182   &  -0.890   &  46.613      &  61.889    &    109.238     &     145.036     \\
 \hline
 110     &   -1.241   &  -0.895  &   64.628     &   89.664    &    156.349      &    216.920      \\
 \hline
 %\hline
\end{tabular}
\end{center}
\end{table}

%%%%%%%%%%%%%%%%%%%%%%%%%%%%%%%%%%%%%%%%%%%%%%%%%%%%%%%%\newpage

\end{document}